\documentclass[twocolumn,a4paper,10pt]{article}
\pdfoutput=1
\usepackage{fullpage}

\usepackage{cite}

\usepackage{times}

\usepackage{graphicx} 
\graphicspath{{graphs/}}
\DeclareGraphicsExtensions{.pdf}

\usepackage{amsmath}
\usepackage{algorithm}
\usepackage{algpseudocode}
\usepackage{array}
\usepackage{mdwmath}
\usepackage{mdwtab}
\usepackage[caption=false, font=footnotesize]{subfig}
\usepackage{fixltx2e}
\usepackage{url}

\usepackage{multirow}
\usepackage{rotating}
\usepackage{paralist}
\usepackage{float}

\usepackage{color}

\definecolor{darkred}{rgb}{0.5,0,0}
\definecolor{darkgreen}{rgb}{0,0.5,0}
\definecolor{darkblue}{rgb}{0,0,0.5}

\usepackage{hyperref}

\hypersetup{ colorlinks,
linkcolor=darkblue,
filecolor=darkgreen,
urlcolor=darkred,
citecolor=darkblue }

\hypersetup{
  pdfauthor={Marcin Nagy, Varun Singh, Joerg Ott, and Lars Eggert},
  pdftitle={Congestion Control using FEC for Conversational Multimedia Communication},
  pdfsubject={RTP Congestion Control using FEC},
  pdfkeywords={RTP} {Congestion Control} {FEC} {RRTCC} {C-NADU} {RMCAT} {WebRTC}
}

\begin{document}
\pagestyle{empty}
%
%
% --- Author Metadata here --- VS:
% \conferenceinfo{WOODSTOCK}{'97 El Paso, Texas USA}
%\CopyrightYear{2007} % Allows default copyright year (20XX) to be over-ridden - IF NEED BE.
%\crdata{0-12345-67-8/90/01}  % Allows default copyright data (0-89791-88-6/97/05) to be over-ridden - IF NEED BE.
% --- End of Author Metadata ---

%\title{Using FEC for Rate Control\\in Conversational Video Communication}
\title{Congestion Control using FEC for Conversational Multimedia Communication}

\author{ Marcin Nagy \\Aalto University     \\marcin.nagy@aalto.fi
  \and Varun Singh \\Aalto University     \\varun.singh@aalto.fi
  \and J\"org Ott   \\Aalto University     \\jorg.ott@aalto.fi
  \and Lars Eggert\\NetApp\\lars@netapp.com
}
\date{5 October 2012}

\maketitle

\begin{abstract}
%\boldmath
%Explain why use FEC for rate control. Claim results.

%As more users are gaining access to broadband Internet, the deployment
%of video communication is increasing. While there are no standardized
%rate-control algorithms for conversational video, most use loss and/or
%delay as indicators to perform congestion control. 

In this paper, we propose a new rate control algorithm for conversational
multimedia flows. In our approach, along with Real-time Transport Protocol
(RTP) media packets, we propose sending redundant packets to probe for
available bandwidth. These redundant packets are Forward Error Correction
(FEC) encoded RTP packets. A straightforward interpretation is that if no
losses occur, the sender can increase the sending rate to include the FEC bit
rate, and in the case of losses due to congestion the redundant packets help
in recovering the lost packets. We also show that by varying the FEC bit rate,
the sender is able to conservatively or aggressively probe for available
bandwidth. We evaluate our FEC-based Rate Adaptation (FBRA) algorithm in a
network simulator and in the real-world and compare it to other congestion
control algorithms.

\end{abstract}

%\category{C.2.2}{Network Protocols}{Applications (SMTP, FTP, etc.)}
%\terms{Algorithms, Design, Performance, Experimentation}
%\keywords{FEC, RTP, Multimedia, Congestion Control}

\section{Introduction}
%- What is FEC and how is it normally used

%- What is the *motivation* of using FEC for Rate-control or What may be the gains.

The development of Web Real-Time Communication (WebRTC) and telepresence
systems is going to encourage wide-scale deployment of video communication on
the Internet. The main reason is the shift from desktop or native real-time
communication (RTC) applications (e.g., Skype, Google Talk, Yahoo and MSN
messenger) to RTC-based web browsers or web applications. Currently, each web
application implements their RTC stack as a plugin, which the user downloads.
With WebRTC the multimedia stack is built into the web-browser internals and
the developers need to just use the appropriate HTML5 API. 

With the expected increase in multimedia traffic, congestion control is a
re-emergent problem. These flows are subject to the fluctuations in path
properties, such as packet loss, queuing delay, path changes, etc. Moreover,
buffer bloat~\cite{gettys:2012:bufferbloat} and drop-tail routers can cause
delay and bursty loss, which affects the user experience. Unlike in elastic
applications, there are normally bit rate constraints on codecs, i.e., the
codec encoding rates have a limited range of adaptation, and can only pick a
few rates between these limits. Moreover, varying the encoding rate too often
or in large steps degrades video quality~\cite{Zink03subjectiveimpression}.
Conversational multimedia differs from streaming multimedia because the former
imposes a strict delay on end-to-end packet delivery. Consequently, it affects
the size of the playout buffer, which for the conversational multimedia flows
is at least an order of magnitude smaller than for streaming.

To tackle congestion control, the IETF has chartered a new working group,
RMCAT\footnote{http://tools.ietf.org/wg/rmcat/}, to standardize
congestion-control for real-time communication, which is expected to be a
multi-year process~\cite{jennings:2013:webrtc}. Therefore, \cite{draft.rtp.cb}
proposes minimum circuit breaker conditions under which a conversational
multimedia flow should terminate its session. While this is not enough to
perform congestion control, it reduces the effect of an unadaptive multimedia
flow on the network~\cite{pv-2013-cb}.

%The popularity of conversational video applications, like VVoIP, has grown
%rapidly in recent years. Despite this rapid increase no satisfactory
%congestion control mechanisms for such applications has been developed.
%Thus, there is a need to create useful mechanisms that are able to quickly
%adapt end-points sending rates to the changes in the network conditions.

%The rate adaptation of multimedia streams is more difficult than of TCP
%flows, as packet losses contribute to degradation of the user experience.

\begin{figure}[!t]
	\centering
	\includegraphics[width=0.8\columnwidth,clip=true, trim=0 1.5cm 0 1cm]{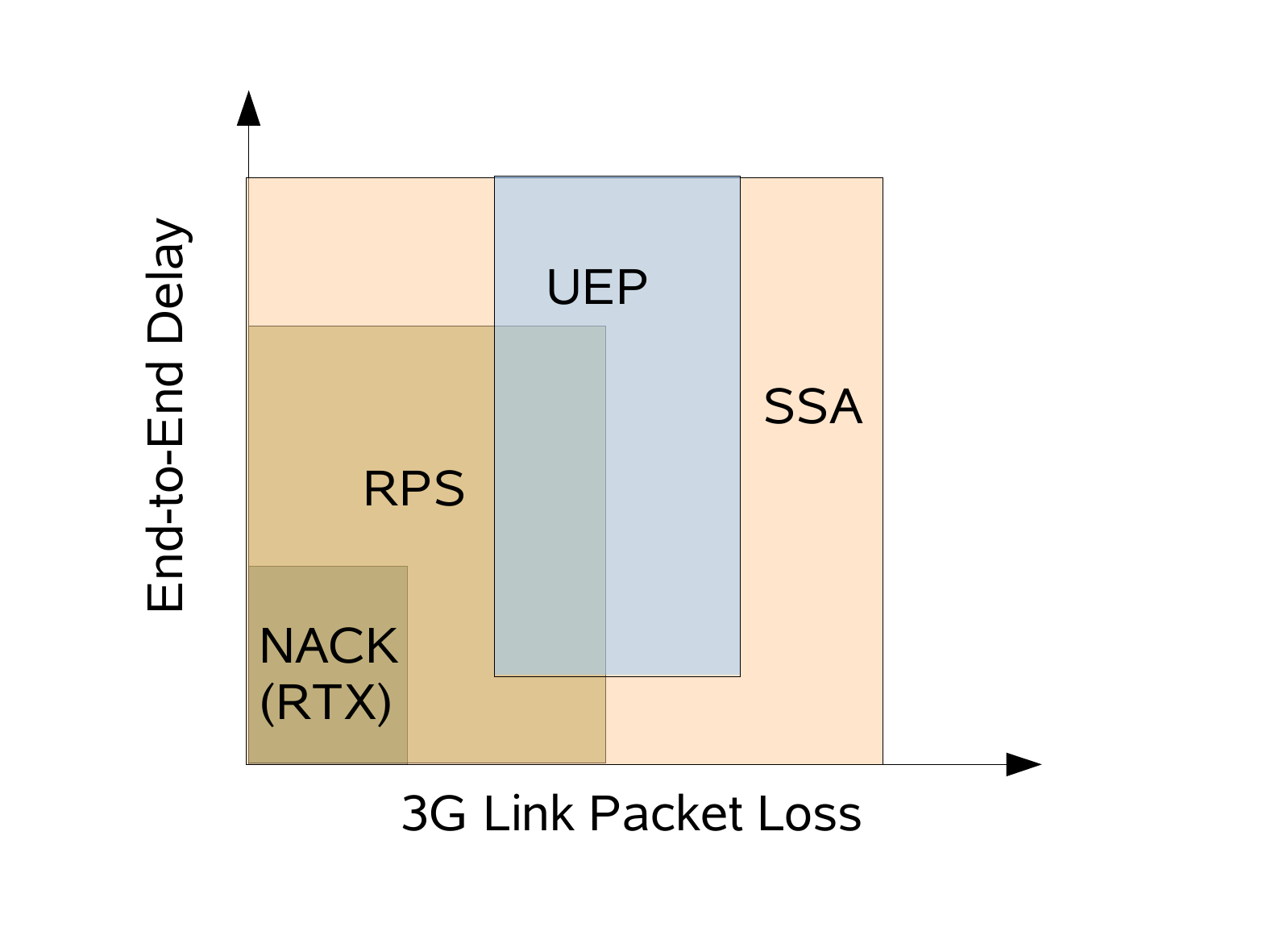}
    \caption{shows the applicability of an error-resilience scheme based on
    the network delay and packet loss~\cite{Devadoss2008}. UEP is Unequal level
    of protection implemented by FEC, SSA is slice size adaptation, RPS is 
    reference picture selection, NACK are retransmissions.}
	\label{apply_err}
\end{figure}

Conversational multimedia systems use Forward Error Correction (FEC) and
Packet Loss Indication (PLI) to protect against packet loss~\cite{664283,
855913, Devadoss2008}, i.e., the endpoint trades-off part of the sending rate
for redundant packets or retransmissions to reduce the effect of losses on the
user experience. An endpoint prefers to use FEC in networks where
retransmissions would arrive later than the playback time of the packet and
the loss rate is deterministic. Figure~\ref{apply_err} shows the applicability
of different error-resilience schemes based on network latency and packet
loss~\cite{Devadoss2008}. If the loss rate differs significantly over several
FEC intervals from the expected rate, the endpoint needs to adapt the FEC rate. To summarize, an
RTC endpoint needs to adapts the sending rate to best fit the changing network
capacity and the amount of FEC (or FEC interval) to best fit the observed
network loss rate. Therefore, in this paper, we investigate the use of
redundant packets not only for error-concealment but also as a probing
mechanism for congestion control (ramping up the sending rate). Other
error-resilience methods like, retransmissions (such as,
ARQ~\cite{Devadoss2008}), sending duplicate packets~\cite{5953578} (reduces
network efficiency), error-concealment at the sender and
receiver~\cite{855913} are ignored.

By choosing a high FEC rate, an endpoint aggressively probes for available
capacity and conversely by choosing a low FEC rate, the endpoint is
conservative in probing for additional capacity. If during probing for additional
capacity, a packet is lost due to congestion, the receiver may be able to
recover it from the FEC packet (i.e., if the FEC arrives in time for decoding).
If no packet is lost, the sender can increase the media encoding rate to
include the FEC rate. This method can be especially useful when the sending
rate is close to the bottleneck link rate, by choosing an appropriate FEC rate
the endpoint can probe for available capacity without affecting the baseline
media stream.

%Issue of adapting the sending rate with minimizing possible losses can be
%resolved by adding redundant data to the multimedia stream to probe the
%network bandwidth. We investigate possibility of using FEC for rate
%adaptation and error recovery simultaneously. By taking this approach it is
%possible to apply an aggressive rate adaptation algorithm, as possible
%losses caused by wrong network conditions judgement during congestion
%period may be recovered using FEC packets, and the FEC stream can be used
%to probe the bandwidth.

To verify that FEC can be applied to multimedia congestion control (for
WebRTC), we propose a new RTP rate adaptation algorithm, termed \textit{FEC
Based Rate Adaptation}--FBRA (see Section~\ref{sec.algorithm}). We evaluate
its applicability using ns-2~\cite{ns2} in various simulation scenarios and in
a real-world testbed (Section~\ref{sec.eval}). Furthermore, to better
understand the concept, we have developed a simplified state machine of the
algorithm (Section~\ref{sec.theory}). Finally, we present also a detailed description
of the design of the \emph{Adaptive Multimedia Subsystem} (Section~\ref{sec.system}).

\section{Related Work}
\label{sec.rel-work}

The Real-time Transport Protocol (RTP)~\cite{rfc3550} is used to deliver
conversational multimedia flows and is favored over TCP for media delivery due
to the very stringent timing requirements of multimedia~\cite{tcp-real-time}.
RTP carries the media packets while the RTCP reports carry media playout and
sender-to-receiver path statistics, such as, jitter, RTT, loss rate, etc. The
media playout information can be used to synchronize the audio and video
streams at the receiver and the path statistics are used by the sender to
monitor the session.

%While standard RTP allows using $5\%$ of the media rate
%for sending RTCP feedback, 
%

Standard RTP limits the reporting interval to a minimum of $5\pm2.5s$ to avoid
frequent reporting. Due to the long reporting interval, the end-to-end path
statistics become too coarse-grained to be applicable for congestion control.
However, RFC4585~\cite{rfc4585} removes this minimum reporting interval
constraint and endpoints can use up to $5\%$ of the media rate for RTCP. With
the smaller reporting interval, the congestion control algorithm can expect
feedback packets on a per-packet, per-frame or per-RTT
basis~\cite{draft.rmcat.feedback}. Using RTCP Extended Reports
(XR)~\cite{rfc3611}, an endpoint can report other path heuristics, such as
discarded packets, bursts and gaps of losses, playout delay, packet delay
variation, among others.

%While there are no standardized rate-control for video communication, several
%have been proposed in the last decade. 

Several sender-driven congestion control algorithms have been proposed over
the years. Most prominent is the TCP Friendly Rate Control
(TFRC)~\cite{tfrc_347397}, which can be implemented using the information
contained in standard RTCP reports (e.g., RTT and loss measurements). However,
TFRC requires feedback on a per-packet basis~\cite{draft.rtp.tfrc}, which can
produce an increase and decrease in the media rate (sawtooth) in a very short
interval of time~\cite{saurin:2006:thesis, VS:Rate}. RAP~\cite{752152} uses a
token bucket approach to additively increase and multiplicatively decrease the
rate (AIMD), but as the media rate reaches the bottleneck rate the encoding
rate in this case as well exhibits a sawtooth-type of behavior. Due to the
impact on perceived media quality, any algorithm that consistently produces a
sawtooth-type of behavior is not well suited for conversational multimedia
communication \cite{Gharai:2002wt,VladBalan:2007dq}.

Instead of just relying on RTT and loss for congestion control,
Garudadri~\textit{et al.}\cite{4397059} also use the receiver playout buffer
to detect under and overuse and schedule RTCP feedback packets every
$200$-$380ms$ to have timely feedback. Singh~\textit{et al.}\cite{VS:HetRate}
use a combination of congestion indicators: frame inter-arrival time, playout
buffer size for congestion control. Zhu~\textit{et al.}\cite{rmcat-nada} use
ECN and loss rate to get an accurate estimate of losses for congestion
control. O'Hanlon~\textit{et al.}\cite{rmcat-dflow} use a delay-based estimate
when competing with similar traffic and use a windowed-approach when competing
with TCP-type cross traffic, they switch modes by using a threshold on the
observed end-to-end delay, their idea is similar to the one discussed
in~\cite{budzisz2011fair}. Holmer~\textit{et al.}\cite{googleRC} proposes a
Receive-side Real-time Congestion Control (RRTCC) algorithm, which uses the
variation in frame inter-arrival time to detect link under and overuse. The
new media rate is calculated at $1s$ intervals by the receiver and signaled to
the sender in a Temporary Maximum Media Stream Bit Rate Request (TMMBR)
message~\cite{rfc5104}. RRTCC also does not react to losses less than $2\%$,
instead increases the rate until 10\% losses are observed. Recent analysis of
RRTCC shows that it performs poorly when competing with
cross-traffic~\cite{fhm-2013-gcc, pv-2013-rrtcc}.

Most of the above literature for congestion-control of conversational
multimedia considers using error resilience and congestion control separately.
Zhu~\textit{et al.}\cite{Zhu:2001tu,springerlink:10.1023/A:1022865704606}
propose using Unequal Loss Protection (ULP) for both congestion control and
error-resilience. Firstly, they estimate the available rate using a variant of
TFRC, called Multimedia Streaming TCP-friendly protocol (MSTFP)~\cite{871542}.
Secondly, they take packet loss and historical sending rate to smooth out the
encoding rate. Lastly, they apply FEC while performing congestion control and
their results show a significant increase in Peak Signal-to-Noise Ratio
(PSNR). MSTFP does not use RTP/RTCP, applies it to streaming video and
acknowledges each packet for calculating the TFRC estimate. While our proposal
applies to conversational video with tight delay requirements.

%In this paper, we use single-error detecting and correcting Parity FEC instead
%of the more sophisticated ULP schemes.

%[VS TODO:] We need to say something about using single-error detecting
%and correcting Parity FEC

%Furthermore, our
%research reveals that due to the additional protection ensured by FEC, 
%the GOP size can be increased which reduces the complexity at the encoder and 
%may even reduce the power consumption of the device.

%Therefore, we believe that the rate adaptation efficiency is significantly
%improved if it is joined with FEC.

%\section{Related Work}
%\label{sec.rel-work}
%
%- Cite congestion control
%
%- Cite prior art for using FEC for congestion control
%
%Note: Related work could also go at the last before conclusions

%%%%%%%%%%%%%%%%%%%%%%%%%

\section{Concept: Using FEC for Congestion Control}
\label{sec.theory}

%- what are the things to consider/challenges etc

%- always overhead needed in rate adaptation

%- theoretical advantages of using FEC

%- should FEC be used constantly or switched on-off

%- FEC overhead

%- what about reduction during losses, should it be switched off??

%-state machine for theoretical model

%- FEC interval adaptation

%- FEC scheme

%- conveying FEC information in RTCP

%- subsection on metrics

%- NS-2 results for constant link rate.

\begin{figure}[!t]
	\centering
	\includegraphics[width=\columnwidth]{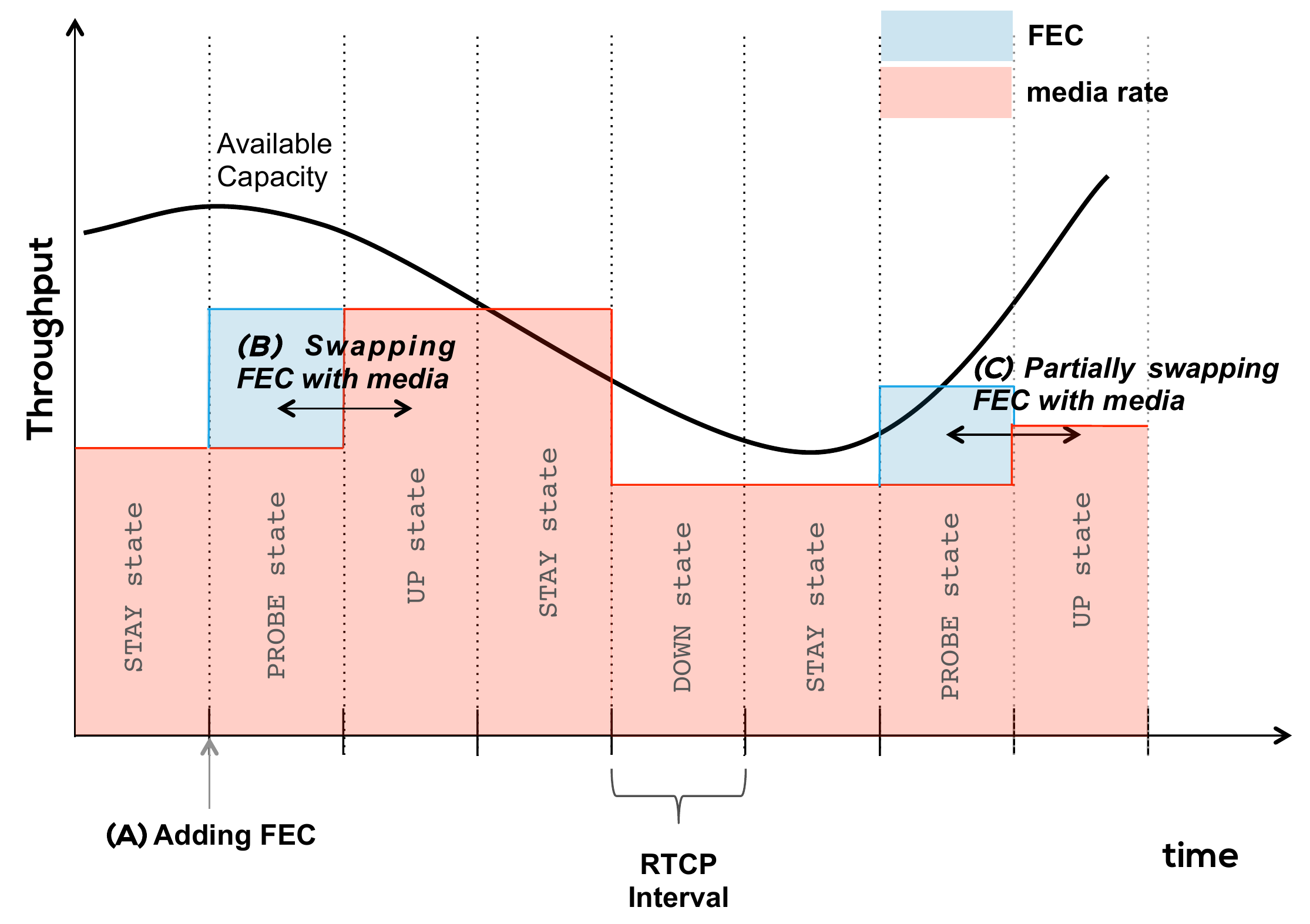}
    \caption{shows the concept of using FEC for rate control. When
    congestion cues indicate no congestion, the FEC stream is enabled to probe
    for available bandwidth. When no losses are reported in the next RTCP
    report, the media bit rate is increased. When congestion is observed, the
    congestion control algorithm reduces the media rate.}
	\label{fec_usage}
\end{figure}

\begin{figure}[!t]
	\centering
	\includegraphics[scale=0.5, trim=0 0.5cm 0 0.5cm]	{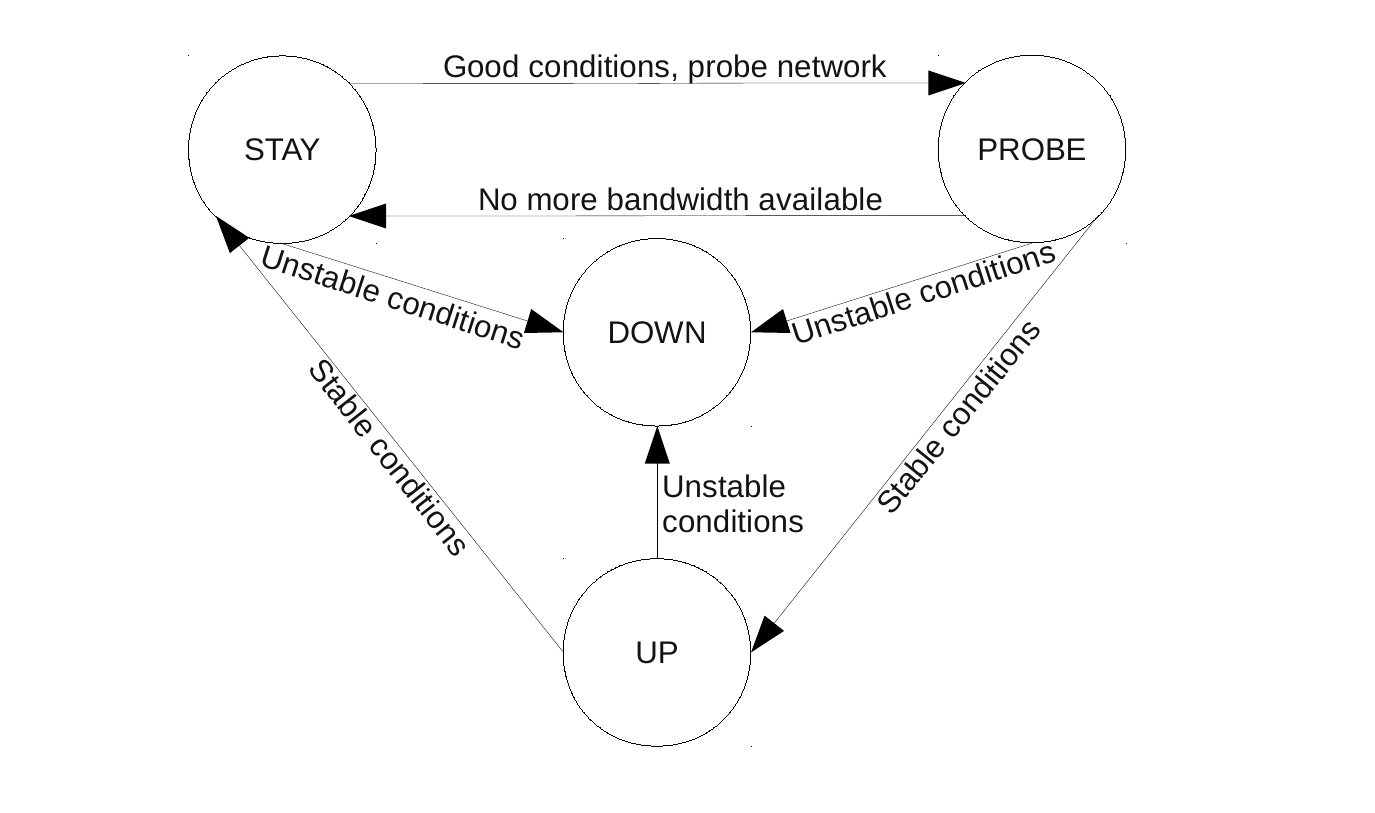}
	\caption{State machine of a congestion control algorithm enabling FEC. 
	It shows that after enabling FEC three conditions may occur. 
	1) No more bandwidth is available, the sender
	should keep the current rate, 2) Stable conditions are detected, the sender 
	should increase the rate and disable FEC, and 3) Unstable conditions are 
	detected, and the sender should reduce the rate to the goodput.}
	\label{model}
\end{figure}

FEC is one method of error protection that improves flow reliability by adding
redundant data to the primary flow which is used by the receiver to recover
parts that have been lost due to either congestion, or bit-errors. The rate
control algorithm on the other hand aims at providing the best possible
network path utilization, but risks over-estimating the available end-to-end
capacity that may lead to congestion induced losses.

The main idea behind using FEC for rate control is to enable FEC alongside the
media stream and use it to probe the path for available bandwidth. If the path
conditions are good and stable after the FEC stream is switched on, the media
encoding rate is increased by the amount of the FEC stream rate and the FEC
stream is disabled (see figure~\ref{fec_usage}). The main advantage of this
approach is that the applied rate control algorithm can be more adventurous in
probing for available capacity, as improved reliability compensates for
possible errors resulting from link overuse.

Figure~\ref{model} illustrates the state machine of a congestion control
algorithm incorporating FEC for probing. The state machine includes 4 states:
\emph{\textbf{STAY}}, \emph{\textbf{PROBE}}, \emph{\textbf{UP}}, and
\emph{\textbf{DOWN}}. The rate adaptation algorithm decides based on the
congestion cues (such as, RTT, loss/discard rate, jitter, packet delay
variation, ECN, etc.) to stay in the current state, or transit to another. The
state machine does not specify the conditions for the transition between
states, but only gives a very generic description of the path conditions and
leaves the interpretation to the underlying congestion control algorithm.

The primary consideration for probing using FEC is \emph{``How much FEC should
be introduced alongside the media stream?''} If the endpoint uses a higher FEC
rate, it has better protection against losses, irrespective of what FEC scheme
it is using. Moreover, the ramp-up is quicker but at the risk of overloading
the path and causing congestion. If the endpoint uses a lower FEC rate, it has
weaker protection against losses and also a potentially slower ramp-up.
Therefore, the sender should observe the congestion cues to make its decision.
If the congestion cues indicate that it is operating close to the bottleneck
link, it should use lower FEC rate; however, if the cues indicate that it is
underutilizing the link it should use a higher FEC rate. Hence, an important
aspect of the rate adaptation algorithm is the ability to find the correct FEC
rate for the current network conditions.
Another aspect to consider is the FEC scheme. It determines
the amount of redundancy injected into the network. 
An application may employ FEC at the packet-level, frame-level or 
use an unequal level of protection (ULP).

\section{FEC-Based Rate-control\\ Algorithm (FBRA)}
\label{sec.algorithm}

%- Discussion on design parameters

%- RTCP interval

%- FEC Interval

%- FBRA Algorithm

%The main idea behind using FEC for rate adaptation and packet recovery simultaneously. By taking this approach it is possible to apply more aggressive rate adaptation algorithm, as possible losses caused by wrong network conditions judgement during congestion period may be recovered using FEC packets. 

%The heart of the rate control in conversational video communication is based on analysis of RTCP reports. However, there are two fundamental issues with rate control in the basic RTP standard \cite{rfc3550}. First, RTCP feedback is sent in at least $5\pm2.5$ second intervals, which limits capabilities of the rate control algorithms to respond to changing network conditions. Second, the RTCP Receiver Report carries only global metrics about reception statistics in the given interval, which are too coarse to effectively assess the congestion status. Thus, to overcome these limitations, we use the extended RTP standard, defined in RFC 4585 \cite{rfc4585}, and broaden content of RTCP reports by adding other metrics from the subset of the RTCP XR standard \cite{rfc3611}.

%Usage of the extended feedback profile \cite{rfc4585} allows to quickly respond to changes in network conditions. For point-to-point sessions, the RTCP transmission interval is almost always dependent on the minimum possible transmission interval, as the 5\% of the session bandwidth is virtually impossible to reach. As the RFC 4585 \cite{rfc4585} standard does not enforce value of the minimum RTCP transmission interval, 

In this section, we describe our proposed congestion control algorithm and the
RTP/RTCP extensions it uses. We also describe the conditions under which to
enable FEC.

\subsection{Using RTP/RTCP Extensions}

Our algorithm uses a short RTCP reporting interval and our experiments 
show that the interval need not be shorter than $2 \times RTT$.
In exigent circumstances, such as the receiver detects that the
playout buffer is about to underflow due to late arrival of 
packets\footnote{Typically underflow may occur due to routing updates 
or queuing delay at an intermediate router.} then the receiver may send 
the RTCP feedback early, however, an endpoint can only 
send an early feedback once every other reporting intervals~\cite{rfc4585}. 

Apart from the congestion cues reported in the standard RTCP 
Receiver Report (RR), such as jitter, RTT, and loss rate, we
additionally use Run-length encoded (RLE) lost~\cite{rfc3611}, 
and RLE discarded~\cite{draft-ietf-xrblock-rtcp-xr-discard-rle-metrics-04.txt} 
packets. Using the RLE lost and discarded packets\footnote{Packets that arrive too late to be displayed are discarded by the receiver.} the sender can 
correlate when exactly the loss and discard events took place in 
the reporting interval. If these events occurred earlier in the reporting interval, 
the sender may ignore them as transient and keep the same sending rate. 
Conversely, if the events occurred later in the interval (more recent) then the sender would 
calculate the exact goodput\footnote{using the history of sent, lost and 
discarded packets in the last reporting interval~\cite{VS:Rate}.} 
and use that as the sending rate. Furthermore, 
using the RLE lost and discarded packet information the receiver 
may be able to distinguish between bit-error losses and congestion
losses~\cite{VS:HetRate}.

RTT, as a congestion cue, has one fundamental issue, namely it assumes that the network paths in
both directions are symmetric and that the congestion upstream is similar to
the congestion downstream, which may not be always
true~\cite{NgamwongwattanaT10}. As a result, we use the technique defined
in~\cite{NgamwongwattanaT10} to precisely calculate the one-way delay (OWD) at
the receiver without the need for clock synchronization and the receiving
endpoint reports the observed OWD in the reporting interval using the
extension defined in~\cite{draft.xr.owd}. Increase in the one-way delay may
indicate a queuing delay in the network, which is an early sign of congestion.
To calculate the uncongested OWD, the sender only collects the reported OWD
from the RTCP RRs that do not report any losses or discards into a data
structure, $OWD_{history}$. The sender then uses the $OWD_{history}$ to
determine the optimum OWD. 

\begin{figure}[!tbp]
	\centering
	\includegraphics[width=0.8\columnwidth, trim=0 2cm 0 2cm]{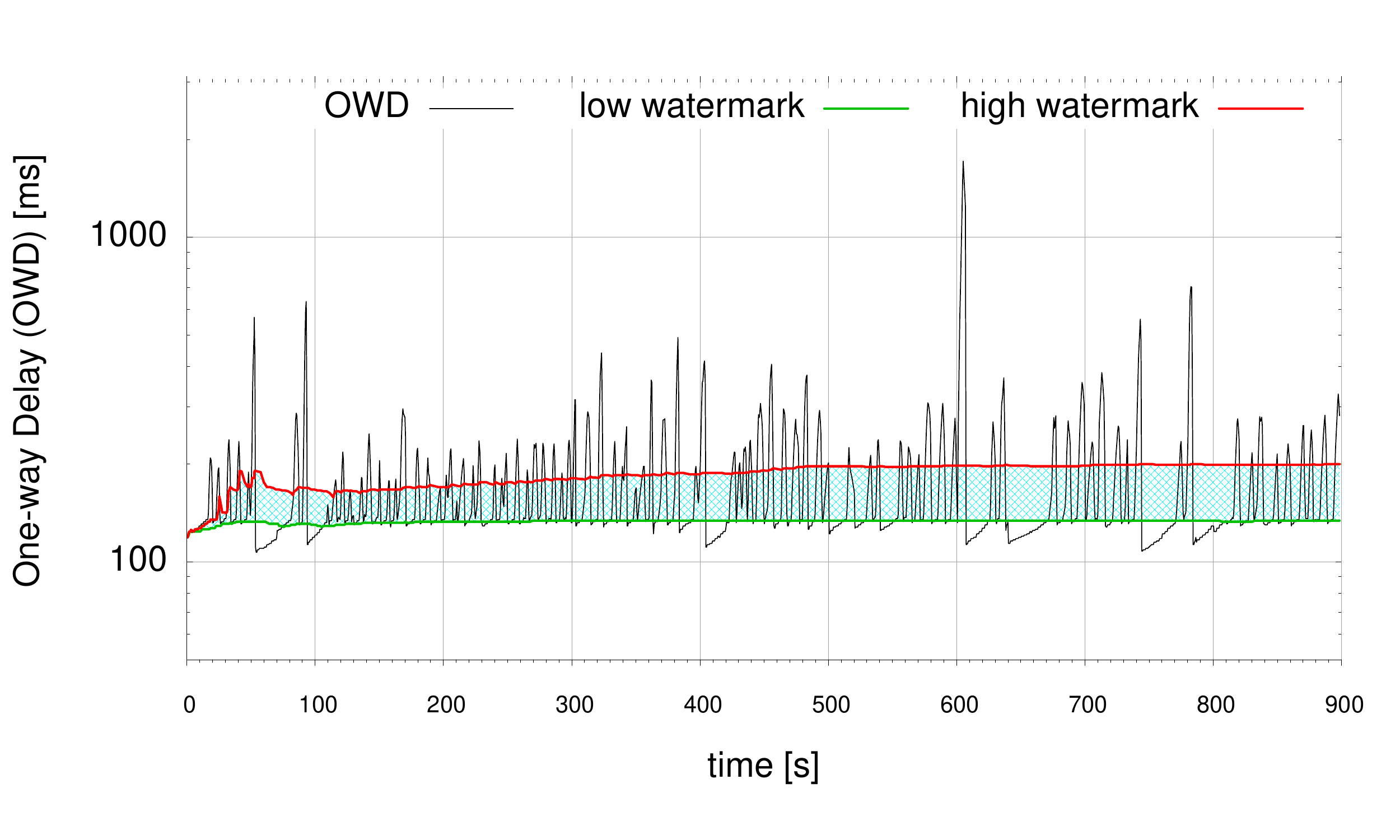}
	\caption{%
	The plot shows the variation in measure OWD value (dark line) during a video call, 
	the low and high watermark represent the $40^{th}$ and $80^{th}$ percentile.
	The FBRA algorithm uses the $\alpha$ and $\beta$ thresholds to validate, if the
	recently measured OWD value falls in the region of operation (shaded area).}
	\label{owd_corrs}
\end{figure}

As most of the $OWD_{history}$ values are close to the ``ideal'' OWD value, we
use the $40^{th}$ percentile\footnote{in our experiments, we find that the
$40^{th}$ percentile value provides better tolerance than the median.} 
to determine the threshold for increasing the sending rate (low watermark),
i.e., if the current OWD measurement is smaller than the $40^{th}$
percentile value indicates link under-utilization.

Similarly, we use the $80^{th}$ percentile, (which is at the higher end of the
distribution) to set the threshold for decreasing the rate (high watermark), 
i.e., if the most recent OWD measurement is larger than the $80^{th}$ percentile 
value indicates the onset of congestion (see Figure~\ref{owd_corrs}). 
This method is very similar to the one used in~\cite{VS:Rate}, 
which uses RTT measurements instead of OWD.

%\small
\begin{align*}
Correlated_{low OWD} = &\frac{current_{OWD}}{40^{th}percentile(OWD_{history})}\\
Correlated_{high OWD} = &\frac{current_{OWD}}{80^{th}percentile(OWD_{history})}
\end{align*}

The FBRA algorithm uses the correlated OWD by checking if its value exceeds thresholds: $\alpha$ and $\beta$.
%in the following way: If the
%$Correlated_{low OWD}$ is lower than a given threshold $\beta$, the
%sending rate is increased, whereas, if the $Correlated_{high OWD}$ is above a
%second threshold $\alpha$, the rate is reduced. 
$\alpha$ and $\beta$ thresholds values must be within [1;2] interval, and have been derived
empirically in a series of experiments. Because these values depend on the
state of the congestion control algorithm, the exact thresholds are described
in more detail in Section~\ref{subsec.fbra}.

%Large $\alpha_{n}$ value  for high OWD makes 
%the FBRA algorithm more aggressive, while for low 
%OWD the algorithm is more conservative. Similarly for $\beta_{n}$.

\subsection{Size of $FEC_{interval}$}
%As this work has concerned usage of FEC for rate control, we have not investigated various FEC schemes, but have decided to use basic per packet scheme. 

%The main concept behind the rate control is the exchange of FEC bandwidth for additional RTP rate if the network conditions are satisfying. Therefore, the amount of added RTP rate has to be variable. We make the FEC rate adaptable by changing the number of RTP packets protected by one FEC packet. We call the amount of protected RTP packets the FEC interval.

In this paper, we use FEC for both rate adaptation and for the error
protection because we can adapt the rate without the fear of creating packet
losses that impair QoE.  We use a basic per packet parity FEC scheme because
it gives $1$ redundant packet for every \emph{N} RTP packets.  By varying the
number of RTP packets  encoded by a single FEC packet  (known as the
$FEC_{interval}$) the congestion control algorithm  can vary the ramp-up rate,
i.e., to increase redundancy the  congestion control algorithm only needs to
reduce the $FEC_{interval}$.  Therefore, for the rest of the paper, we have
ignored the  use of more complex FEC  schemes and leave it for further study.

%\textbf{Size of} $\mathbf{FEC_{interval}}$: 
We have limited the $FEC_{interval}$ to encode between $2$ and $14$ RTP
packets. The minimum $FEC_{interval}=2$ is the lowest possible and creates
maximum redundancy and high FEC rate, whereas, the maximum $FEC_{interval}=14$
creates low redundancy and a low FEC rate. In our experiment, we found that a
$FEC_{interval}>14$ had negligible impact on both congestion control (because
the additional FEC rate is too small) and for error protection (because the
FEC packet is generated too late to help in decoding the lost packet, the
playout buffer of an interactive flow is very small).

%concluded that benefit from adding smaller amount of rate if the FEC interval is higher than 14 is negligible, whereas the protection against losses decreases if the interval is larger. 

The $FEC_{interval}$ is calculated based on the following observations. When
the RTP media rate is low, it is assumed that there is a lot of available
end-to-end capacity for the media stream, and the $FEC_{interval}$ is set low.
This allows the sender to quickly ramp-up the sending rate, and if the link is
overused it also provides higher protection against possible losses.
Conversely, when the RTP media rate is high, it is assumed that the sender is
approaching the bottleneck link capacity and there may not be much more
bandwidth left for the media stream, and the $FEC_{interval}$ is set to a high
value. This allows the sender to ramp-up slowly and avoid overshooting the
bottleneck link capacity. To determine if the current media rate is $low$ or
$high$, the sender keeps a history of all the rates (goodput, sending rate,
combined FEC and RTP sending rate) recorded in the last two seconds, and
compares the current rate with the highest recorded value in the set and with
the initial goodput.

%\begin{figure}[h]
%	\centering
%	\includegraphics[scale=0.5]{graphs/fbra-states.eps}
%	\caption{FBRA state machine}
%	\label{fbra-states}
%\end{figure}

\subsection{FBRA: Algorithm}
\label{subsec.fbra}

The FBRA algorithm has the following states: \emph{STAY}, \emph{PROBE},
\emph{UP}, \emph{DOWN}. The state names match exactly the ones described in
Section~\ref{sec.theory}.
The algorithm changes state depending on the 
information received in the new RTCP RR %(See Algorithm~\ref{fbra:main}) 
and is constrained by the following state transition rules. 

%\begin{algorithm}[!htbp]
%\caption{FEC-Based Rate Adaptation algorithm (FBRA)}
%\label{fbra:main}
%{\fontsize{8}{8}\selectfont
%\begin{algorithmic}[1]
%\If{$Disabled\ Rate\ Control$}
%	\State $NewState \gets s-$
%\ElsIf {$Early\ feedback$}
%	\State $Undershoot\ and\ disable\ rate\ control$
%	\State $Disable\ FEC$
%\ElsIf{$CurrentState=s+$}
%	\State \Call{StateS+}{}
%\ElsIf{$CurrentState=u$}
%	\State \Call{StateU}{}
%\ElsIf{$CurrentState=d$}
%	\State \Call{StateD}{}
%\Else
%	\State \Call{StateS-}{}
%\EndIf
%\end{algorithmic}
%}
%\end{algorithm}

In the \textbf{\emph{DOWN} state} the algorithm (see
Sub-algorithm~\ref{fbra:d}), reduces the sending rate by executing the
\emph{undershooting procedure}. If no congestion is reported in the next RTCP
interval the algorithm transits to the \emph{STAY} state. However, if losses and discards still appear the algorithm stays in the \emph{DOWN} state. In the edge case, when high OWD values are reported, the \emph{DOWN} state is also kept. As the algorithm should not be very sensitive to high OWD values during congestion, the threshold for the edge case is set to $\alpha_{undershoot}=2.0$.
%
%As the \emph{DOWN} state is entered during
%congestion, high OWD values are expected to be reported, and thus the
%algorithm should not be very sensitive this measurement. Therefore we have set
%the $\alpha_4$ parameter value to 2.0. (See Sub-algorithm~\ref{fbra:d} for
%implementation details)

%\input{fbra}
\floatname{algorithm}{Sub-algorithm}

\begin{algorithm}[!tbp]
\renewcommand{\thealgorithm}{1a}
\caption{DOWN state function}
\label{fbra:d}
{\fontsize{8}{8}\selectfont
\begin{algorithmic}[1]
\Function{State DOWN}{}
	\If{$Recent\ Losses \ensuremath{\lor} Discards$}
		\If{$PreviousState=DOWN$}
			\State $NewState \gets STAY$
		\Else
			\If{$Discards \ensuremath{\land} No\ losses$}
				\State $Undershoot\ without\ disabling\ rate\ control$
			\Else
				\State $Undershoot\ and\ disable\ rate\ control$
			\EndIf
			\State $NewState \gets DOWN$
		\EndIf
	\ElsIf{$Corr_{highOWD}>\alpha_{undershoot}$} \Comment{\ensuremath{\alpha_{undershoot}=2.0}}
		\State $Undershoot\ and\ disable\ rate\ control$
		\State $NewState \gets DOWN$
	\Else
		\State $NewState \gets STAY$
	\EndIf
	\State $Disable\ FEC$
\EndFunction
\end{algorithmic}
}
\end{algorithm}

In the \textbf{\emph{STAY} state} the algorithm (see
Sub-algorithm~\ref{fbra:sminus}), keeps the sending rate constant, and the FEC
packets are not generated. The algorithm can remain in this state and not
probe for additional capacity if the congestion cues indicate that the sender
is operating very close to the bottleneck link capacity. Otherwise, it can
transit to the \emph{PROBE} state for probing the path for additional
capacity or for error-resilience. In either case, the stream may benefit with
stability or increased error-resilience. In order to switch to the
\emph{PROBE} state, the $Correlated_{high OWD}$ must be lower than
$\alpha_{stay}=1.1$. Direct transition to the \emph{PROBE} state is not possible
when the current sending rate is higher than $90\%$ of the highest rate
recorded in the last 2 seconds. In this case, the algorithm assumes that the
current rate is operating close to the bottleneck link capacity, and it must
make sure that current rate is stable by staying in the current state for one
more RTCP report interval before probing further. Obviously, if congestion is
detected, the algorithm goes from the \emph{STAY} state to the \emph{DOWN}
state.

\begin{algorithm}[!tbp]
\renewcommand{\thealgorithm}{1b}
\caption{STAY state function}
\label{fbra:sminus}
{\fontsize{8}{8}\selectfont
\begin{algorithmic}[1]
\Function{State STAY}{}
	\If{$Losses$}
		\If{$Recent\ Losses$}
			\State $Undershoot\ and\ disable\ rate\ control$
			\State $New State \gets DOWN$
		\Else
			\State $NewState \gets STAY$
		\EndIf
		\State $Disable\ FEC$
	\Else
		\If{$Recent\ Discards$}
			\State $Undershoot,\ disable\ rate\ control\ and\ FEC$
			\State $NewState \gets DOWN$
		\Else
			\If{$Corr_{highOWD}>\alpha_{stay}$} \Comment{\ensuremath{\alpha_{stay}=1.1}}
				\If{$PreviousState=STAY$}
					\State $Undershoot\ and\ disable\ rate\ control$
					\State $NewState \gets DOWN$
				\Else
					\State $NewState \gets STAY$
				\EndIf\State $Disable\ FEC$
			\Else
				\State $NewState \gets PROBE$
				\State $Enable FEC$
			\EndIf
		\EndIf
	\EndIf
\EndFunction
\end{algorithmic}
}
\end{algorithm}

In the \textbf{\emph{PROBE} state} the algorithm (See
Sub-algorithm~\ref{fbra:splus}) maintains the current sending rate, but sends
FEC packets alongside. If the next RTCP report shows signs of congestion, the
algorithm disables FEC and goes back to the \emph{STAY} state and if further
congestion is detected, it goes to the \emph{DOWN} state. Otherwise it
normally transitions to the \emph{UP} state. When no losses and discards are
observed, the state transition is based on the measured OWD. If the observed
OWD is higher than the $Correlated_{high OWD}$, the sender assumes that
congestion is severe and thus cuts the sending rate more drastically. Our experiments 
show that the best results are obtained when $\alpha_{down}$ parameter, which is the threshold 
for entering the \emph{DOWN} state is between $1.4$ and $1.6$. Because in this state, it is desirable to find out if the link is underutilized, less sensitivity is needed. Thus, we choose $\alpha_{down}=1.6$. Less sensitivity, increases possibility of getting discards, but appearance of them allows us to find out about the link limit. Transition back to the \emph{STAY} state occurs for
correlation values exceeding $\alpha_{stay}=1.1$ (similarly to the \emph{STAY} state). Furthermore,
the algorithm may also decrease amount of the FEC rate if the OWD value is
unexpectedly higher. This condition is checked by comparison the
$Correlated_{low OWD}$ value against the $\beta$ parameter. If the OWD value
is low enough the \emph{UP} state is entered, thus $\beta$ parameter should
be close to the lower boundary of [1;2] interval (we use $\beta=1.2$).

\begin{algorithm}[!tbp]
\renewcommand{\thealgorithm}{1c}
\caption{PROBE state function}
\label{fbra:splus}
{\fontsize{8}{8}\selectfont
\begin{algorithmic}[1]
\Function{State PROBE}{}
	\If {$Recent\ Losses \ensuremath{\lor} Recent\ Discards$}
		\State $Undershoot,\ disable\ rate\ control\ and\ FEC$
		\State $New State \gets DOWN$
	\ElsIf{$Losses \ensuremath{\lor} Discards$}
		\State $NewState \gets STAY$
		\State $Disable\ FEC$
	\Else
		\If{$Corr_{highOWD} > \alpha_{down}$}	\Comment{\ensuremath{\alpha_{down} \in [1.4;1.6]}}
			\State $Undershoot,\ disable\ rate\ control\ and\ FEC$
			\State $New State \gets DOWN$
		\ElsIf{$Corr_{highOWD} > \alpha_{stay}$} \Comment{\ensuremath{\alpha_{stay}=1.1}}
			\State $NewState \gets STAY$
			\State $Disable\ FEC$
		\ElsIf{$Corr_{lowOWD} > \beta$} \Comment{\ensuremath{\beta=1.2}}
			\State $Increment FEC interval$
			\State $NewState \gets PROBE$
		\Else
			\State $NewState \gets UP$
			\State $NewRate \gets CurrentRate+FECRate$
			\State $Disable\ FEC$
		\EndIf
	\EndIf
\EndFunction
\end{algorithmic}
}
\end{algorithm}

%It is also possible that the report content indicates an in-between situation (the one-way delay value is higher than expected, but there are no losses and discards). In such a situation, the algorithm either goes back to the \emph{s-} state, or it stays in the current state with the FEC rate reduced.

\begin{algorithm}[!tbp]
\renewcommand{\thealgorithm}{1d}
\caption{UP state function}
\label{fbra:u}
{\fontsize{8}{8}\selectfont
\begin{algorithmic}[1]
\Function{State UP}{}
	\If{($Recent Losses \ensuremath{\lor} Discards \ensuremath{\lor} Corr_{highOWD}>\alpha_{down}$} \Comment{\ensuremath{\alpha_{down} \in [1.4;1.6]}}
		\State $Undershoot\ and\ disable\ rate\ control$
		\State $NewState \gets DOWN$
	\Else
		\State $NewState \gets STAY$
	\State $Disable\ FEC$
	\EndIf
\EndFunction
\end{algorithmic}
}
\end{algorithm}

In the \textbf{\emph{UP} state} the algorithm (See Sub-algorithm~\ref{fbra:u})
increases the sending rate by replacing the FEC rate with additional RTP media
rate. If congestion is detected, the algorithm transits to the \emph{DOWN}
state, or else to maintain stability for one more reporting interval, it
transits to the \emph{STAY} state. Transition to the \emph{DOWN} state happens
when the $Correlated_{high OWD}$ exceeds $\alpha_{down}$ value. We use $\alpha_{down}=1.4$, as in the \emph{UP} state more sensitivity for early congestion indication is required.

%\newpage

The algorithm is also sensitive to the RTCP reporting
interval duration. This means that if the sender receives an RTCP RR at an interval
shorter than $1.5 \times RTT_{median}$, it assumes this is an early report and
immediately transits to the \emph{DOWN} state. Furthermore, If no RTCP report
is received for $2s$, the sender halves the rate entering the
\emph{DOWN} state~\cite{draft.rtp.cb}.

\subsection{FBRA: Undershooting \& Bounce-back Procedure}

We define two additional procedures: the \emph{undershooting} and the
\emph{bounce-back} procedure. The undershooting procedure attempts to reduce
the overuse of the network caused by the stream. The stream sets the new
sending rate to a value lower than the current goodput~\cite{VS:Rate,
googleRC}, i.e., the sender calculates the new sending rate by subtracting
twice the difference between the current sending rate and the current goodput
and taking $90\%$ of the obtained value.

%After that it checks if the new sending rate is at least $75\%$ of the old sending 
%rate and more than the minimum rate of $32kbps$. 
%[VS Note: 32kbps is a simulation default, not FBRA's]

After undershooting, if the sender predicts that the congestion cues in the
upcoming reports may still show signs of congestion due to the previous
overuse, the sender may disable the rate adaptation for a brief period of
time. This action prevents any further reduction in the sending rate, while
waiting for the congestion cues to stabilize. The deactivation period is is a
bit more than the \emph{RTCP Interval}, i.e., the sender ignores the next RTCP
report that arrives. After the deactivation period expires, the bounce-back 
procedure is executed.

In the bounce-back procedure, the sender examines the most recent RTCP report
(the one received after the period expires), and if there are no signs of
congestion, it increases the sending rate to $90\%$ of the goodput stored
during the undershooting. The algorithm in this case attempts to gradually
bring the sending rate back to the goodput observed at the moment of
undershooting.
However, if the new RTCP RRs continues to report congestion, the FBRA once
again enters the undershooting procedure and this time does not disable the
congestion control.

%%%%%%%%%%%%%%%%%%%%%%%%%
\section{Performance Evaluation}
\label{sec.eval}

%- NS2 simulation setup

%- each NS2 scenarios in a sub-section

%- probably a big table of results/few graphs for some *interesting* cases

In this section, we evaluate the performance of our proposed algorithm, FBRA,
RRTCC~\cite{googleRC} and C-NADU~\cite{VS:HetRate} in ns-2~\cite{ns2}. Our
simulations comprise of the following scenarios: a single RTP flow on a
variable link capacity, one or more RTP flows competing on a bottleneck link,
and one or more RTP flows competing with other short TCP flows on a bottleneck
link. This paper mainly focuses on the congestion control in the Internet
environment, where bit-error losses are few, therefore, we use simple
duplex-links with no link loss rate. For intermediate routers we set the queue
length to 50 packets (ns-2 default) and drop-tail queuing strategy, i.e., the
packet loss observed in the simulation results are due to link overuse. In all
the scenarios, we further divide the simulations into three sub-scenarios with
each sub-scenario using a different bottleneck link delay ($50$ ms, $100$ ms,
and $240$ ms)~\cite{tcp-real-time, s4.eval.fw, rmcat.req}. For statistical
relevance, each sub-scenario is simulated 30 times and the standard deviation
is noted for each metric. The simulation scenarios and the corresponding
evaluation parameters are based on those defined in the RTP evaluation
framework~\cite{ott2012evaluating}.

Furthermore, as the simulations are performed at the packet/frame level, we
decided not to differentiate between the different types of video frames
(namely, I-, P-frames), but represent each frame as a packet of equal size
corresponding to the instantaneous sending rate. However, the sender may
fragment large frames (at high bit rate) to fit the MTU size ($=1500$ bytes).
The endpoints generate the frames at $30$ FPS and all three congestion control
algorithms use the same frame rate and packetization methodology. Startup rate
is outside the scope of this paper and therefore both endpoints begin their
session with an initial sending rate of $128 kbps$, and restrict the minimum
rate for each congestion control algorithm to $32kbps$. There is no
restriction on the maximum rate but, the maximum allowed end-to-end packet
delay ($delay_{max}$) for a packet is set to $400ms$~\cite{s4.eval.fw},
packets arriving after this cut-off are discarded by the receiver without
sending them to the playout buffer.

%In the following sub-sections, we analyze the performance of the three 
%rate-control algorithms in the defined scenarios.

\subsection{Metrics}

Apart from the standard metrics---e.g., goodput, Peak signal-to-noise ratio
(PSNR), Packet Loss Rate (PLR), Average Bandwidth Utilization (ABU)---for
evaluating the impact of the congestion control algorithm, we propose three
new metrics to evaluate the performance of using FEC for rate control.

%, they are: 
%FEC Rate-Control Correctness (FRCC), FEC Frame Recovery Efficiency (FFRE), 
%and TCP fair-share (TFS).  

\textbf{FEC Rate-Control Correctness}: FRCC specifies %in percentage value the
times the fraction of time the congestion control algorithm correctly uses
FEC, i.e., the algorithm starts from the \emph{STAY} state, enables FEC and
returns to the \emph{STAY} state without entering the \emph{DOWN} state.
Consequently, the decision is incorrect if the algorithm enters the
\emph{DOWN} state after enabling FEC. Figure \ref{model} shows which state
transitions are allowed and which are not. Formally, we define:

{\small
\begin{equation*}
FRCC = \frac{count(FEC\ raises\ rate)+count(FEC\ keeps\ rate)}{count(FEC\ enabled)}
\end{equation*}
}

\textbf{FEC Frame Recovery Efficiency}: FFRE specifies the fraction of
successfully recovered frames from all the lost frames that were protected by
FEC. Formally:

{\small
\begin{equation*}
FFRE = \frac{frames_{recovered}}{frames_{protected\ but\ lost}+frames_{recovered}}
\end{equation*}
}

\textbf{TCP fair share}: TFS specifies the ratio between the TCP flow
throughput and the fair amount of bandwidth that should be granted to it. The
ratio can be greater than $1$, if TCP uses more than its fair share. Formally:

{\small
\begin{equation*}
TFS = \frac{\left( \frac{TCP\ Throughput}{no. of\ TCP\ flows}\right)}{\left( \frac{Total\ Throughput}{no.\ of\ flows}\right)}
\end{equation*}
}

%\subsection{Variable link capacity scenario}

%In this simulation scenario the dumbbell topology with two RTP nodes
%simulating video conversation was used. These nodes were connected to the
%bottleneck link whose capacity was made variable. Access links had 100Mb/s
%bandwidth available, and their delay was 1ms. The reverse path was
%symmetrical.  Figure \ref{scenario-1} illustrates this scenario topology.

%The bottleneck link capacity varied between 100kb/s and 256kb/s and followed
%the periodic pattern presented in the figure \ref{capacity-changes}. It
%included many types of capacity changes from very rapid ones to smooth ones.
%We believe that such diverse pattern of capacity changes well verifies
%performance of presented rate control algorithms in different situations.

%\begin{figure}[!htbp]
%	\includegraphics[width=\columnwidth]{capacity.png}
%	\caption{Variation in available capacity at a bottleneck link when only one 
%	RTP flow traverses it. In this scenario we evaluate the reactivity and the
%	ramp-up of the rate-control algorithms. 
%	{\color{red}Note: figure should be from 0-300kbps, font bigger and 
%	y/x-axis ratio=0.75, so that the graph looks wider}}
%	\label{capacity-changes}
%\end{figure}

\begin{figure}[!t]
	\subfloat[]{\label{scenario-1}
	\includegraphics[width=\columnwidth, trim=0 3cm 0 2cm]{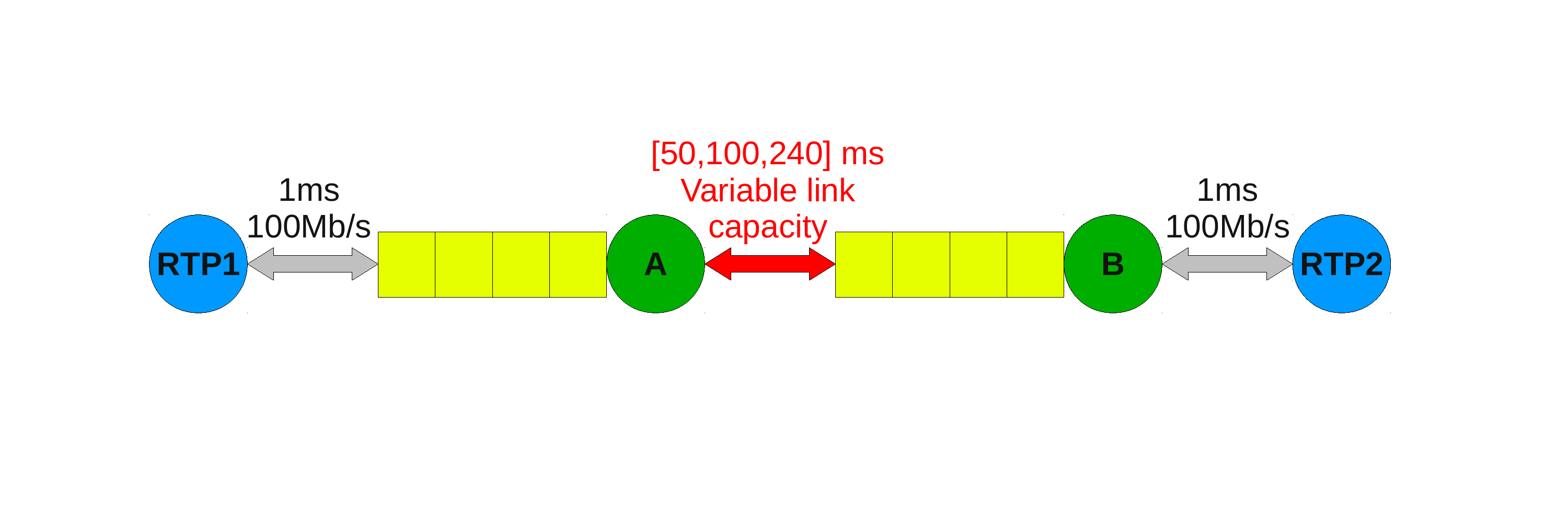}}
	\\
	\subfloat[]{\label{scenario-2}
	\includegraphics[width=\columnwidth]{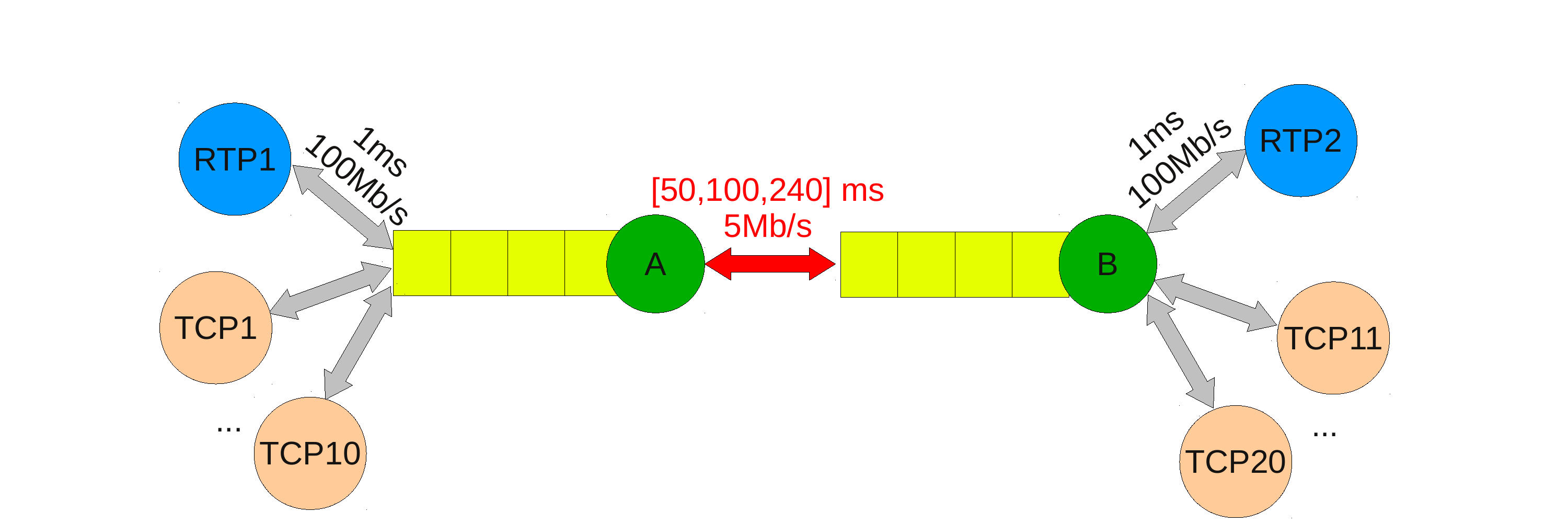}}
	\\
	\subfloat[]{\label{scenario-3}
	\includegraphics[width=\columnwidth, trim=0cm 2cm 0 1cm]{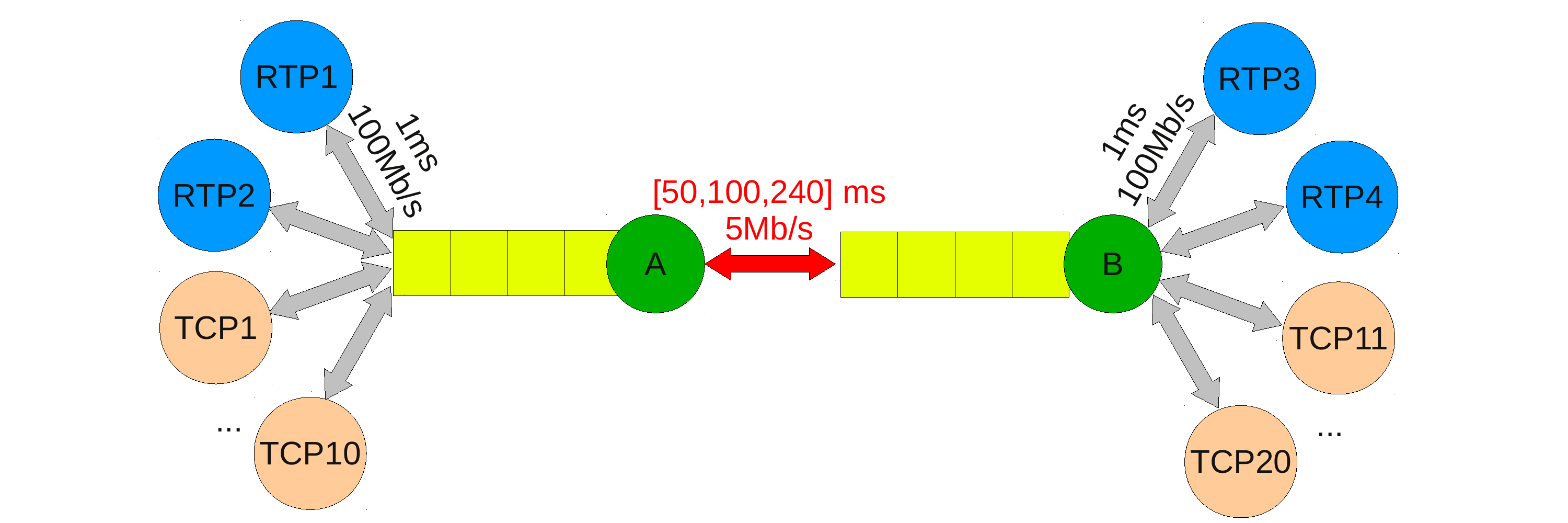}}
	\caption{Shows the \emph{ns-2} simulation topologies for (a) single RTP flow on a variable
	capacity link and (b) One or more RTP flows competing with multiple short TCP flows.}
	\label{ns2-scenarios}
\end{figure}

\begin{figure*}[!t]
	\subfloat[OWD=50ms,  FRCC=92.6\%]{\label{vbr-time-50ms}%
	\includegraphics[scale=0.2]{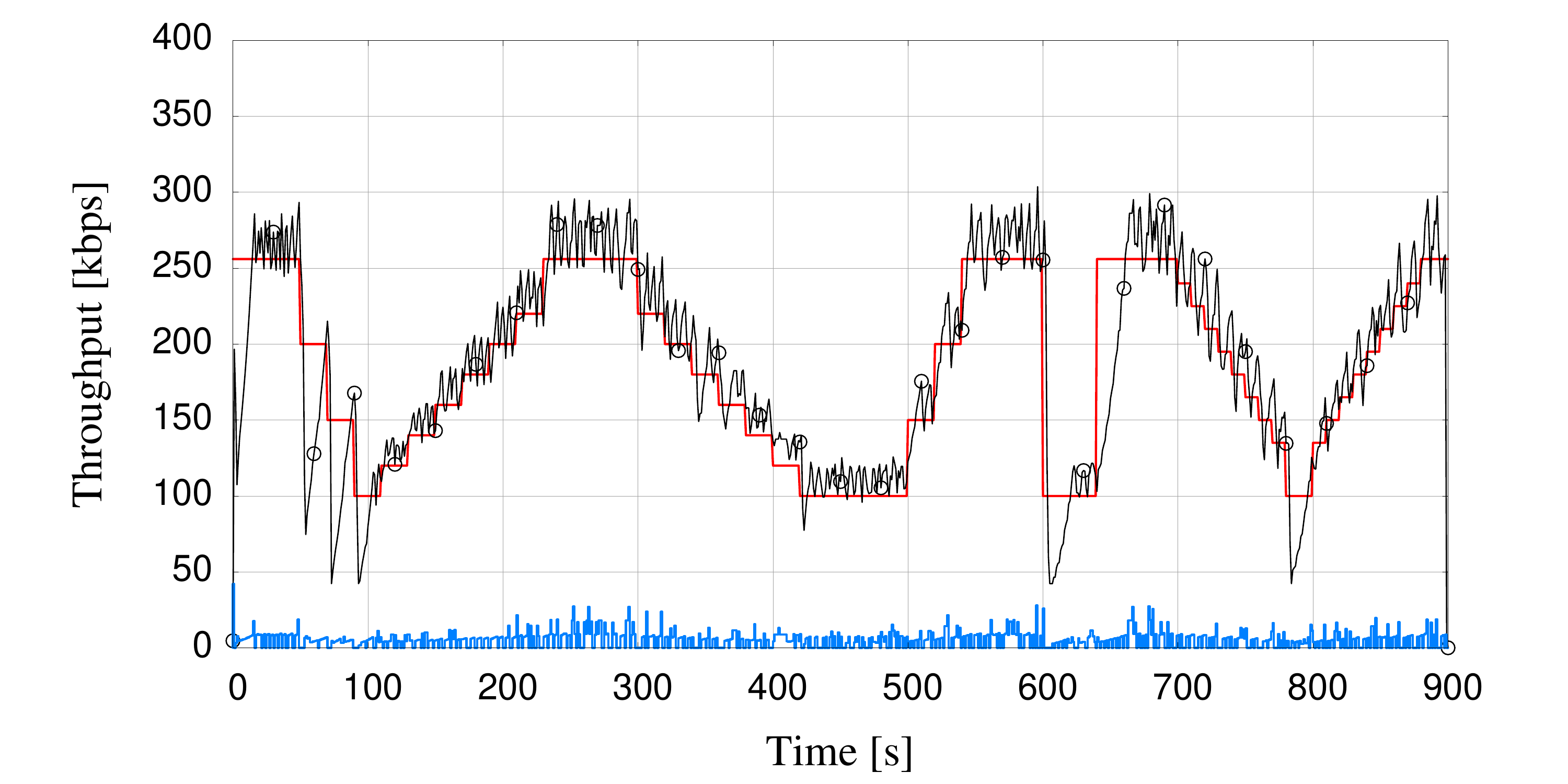}}
	\subfloat[OWD=100ms, FRCC=95.9\%]{\label{vbr-time-100ms}%
	\includegraphics[scale=0.2]{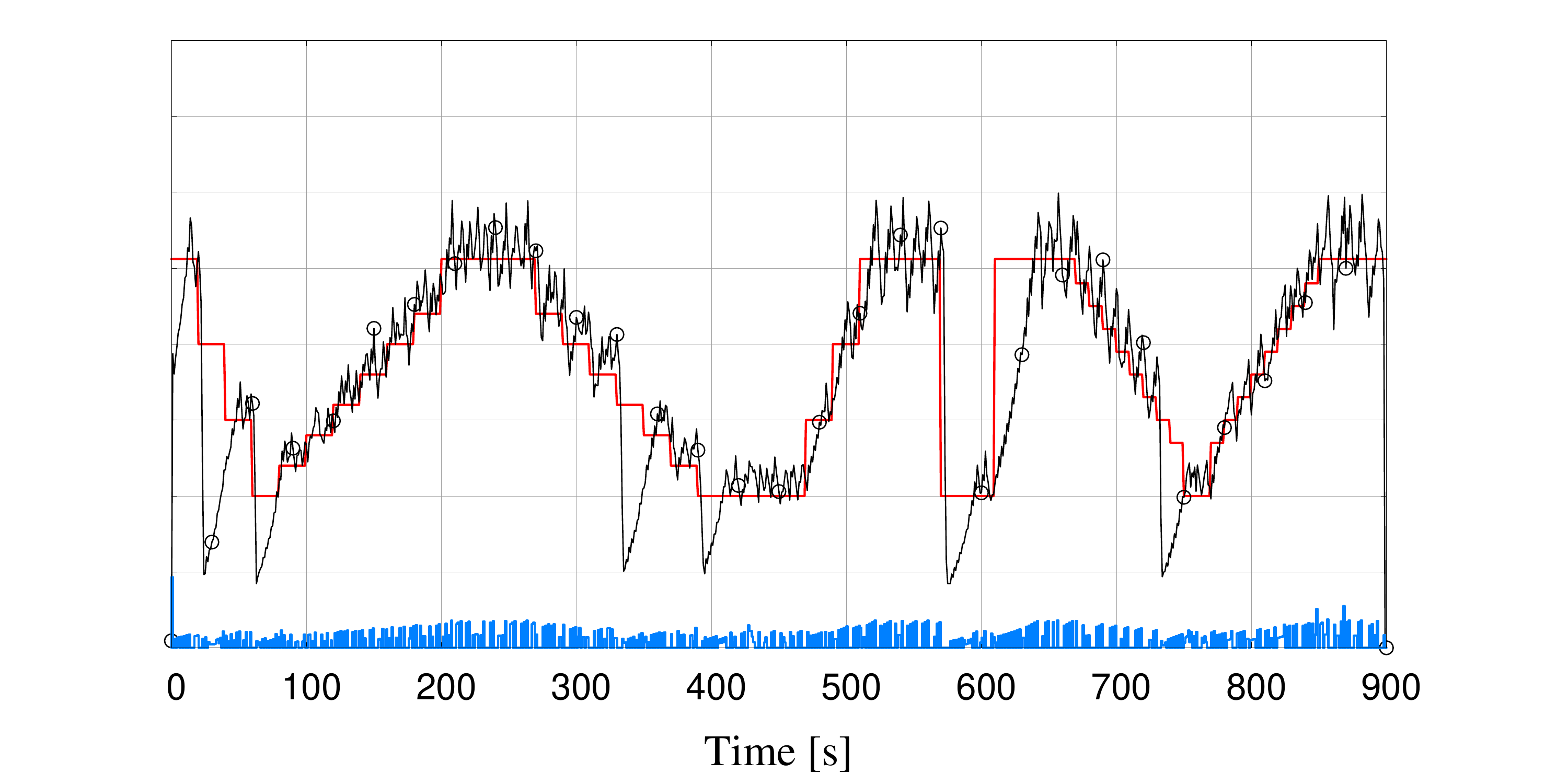}}
	\subfloat[OWD = 240ms, FRCC=90.1\%]{\label{vbr-time-240ms}%
	\includegraphics[scale=0.2]{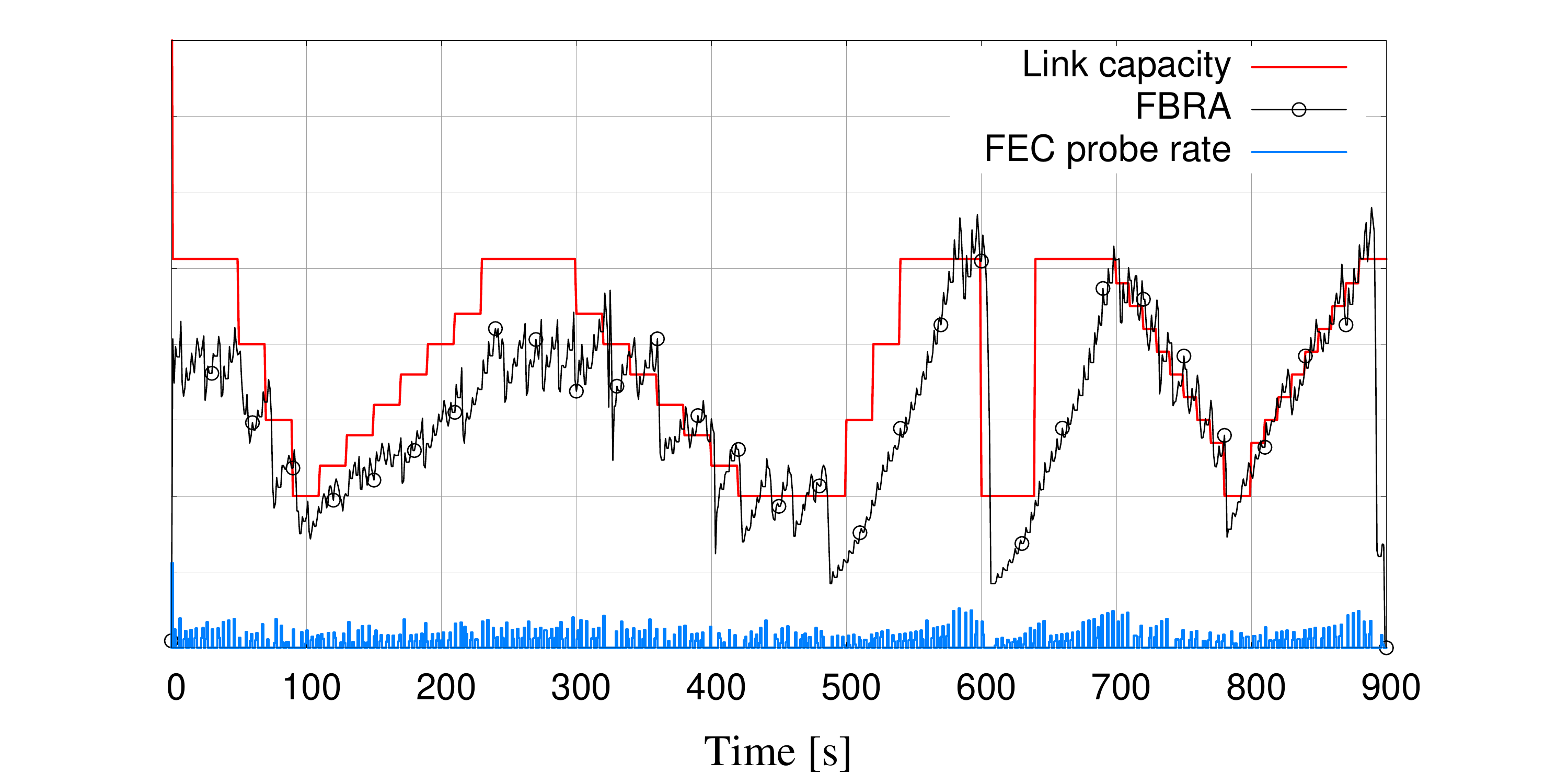}}
	\caption{The plot shows the performance of a single RTP flow using FBRA in a
	varying link capacity scenario with different bottleneck delays. The plots also
	show the FEC probing rate. We observe that the FEC rate is low when the FBRA rate 
	drops and FEC rate is high when the FBRA is ramping-up, but with $FRCC>90\%$ in
	all the cases shows that the FEC probing was accurate. 
	The bandwidth utilization in the high delay  ($240ms$) scenario is low because 
	the FBRA senses  that the one-way delay is  very close to the $delay_{max}(=400ms)$ 
	and is conservative in its bandwidth probing.}
	\label{capacity-changes}
\end{figure*}

\begin{figure*}[!t]
	\subfloat[OWD=50ms,  FRCC=90.5\%]{\label{competition-time-50ms}%
	\includegraphics[scale=0.2]{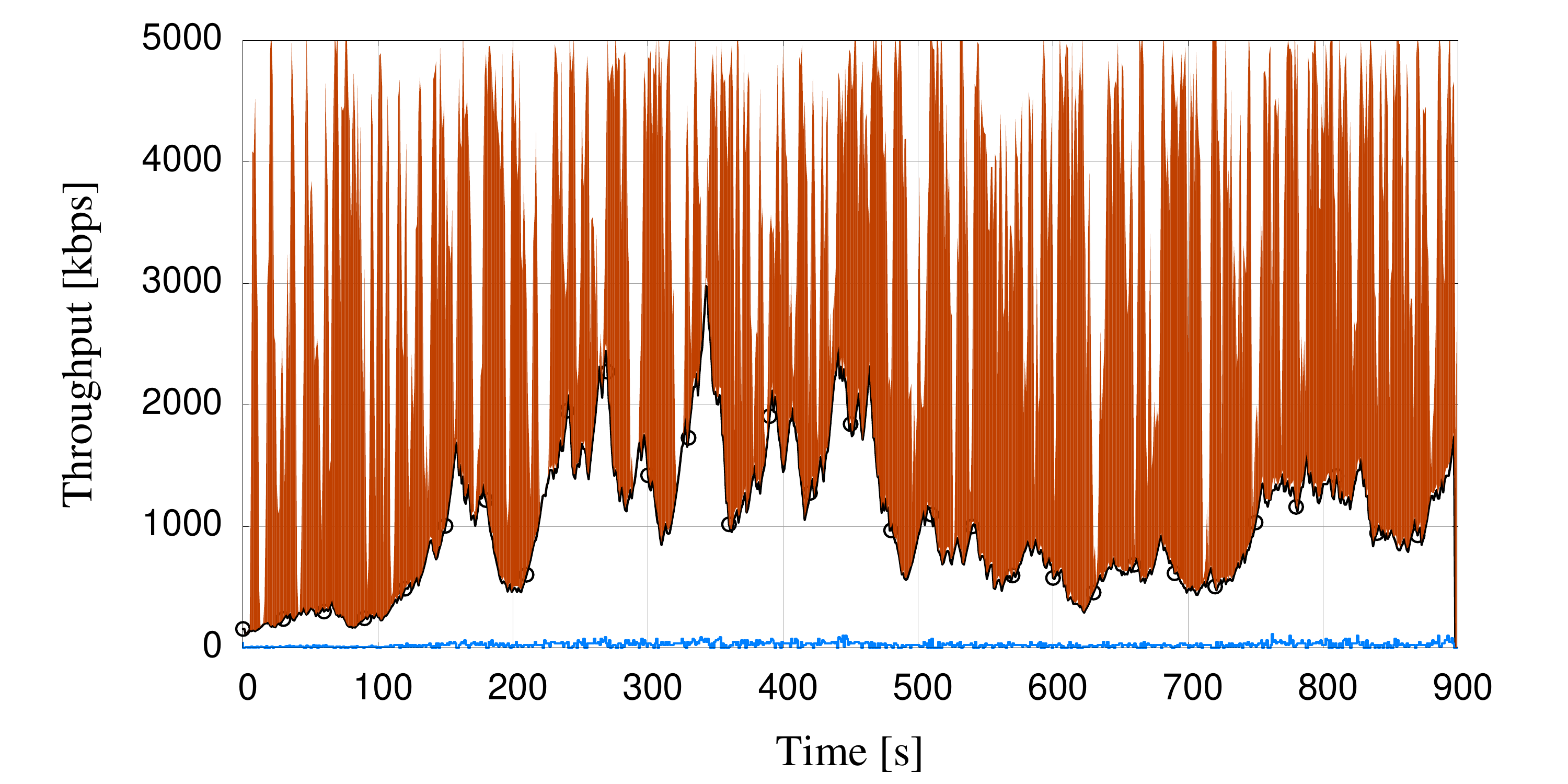}}
	\subfloat[OWD=100ms, FRCC=87.8\%]{\label{competition-time-100ms}%
	\includegraphics[scale=0.2]{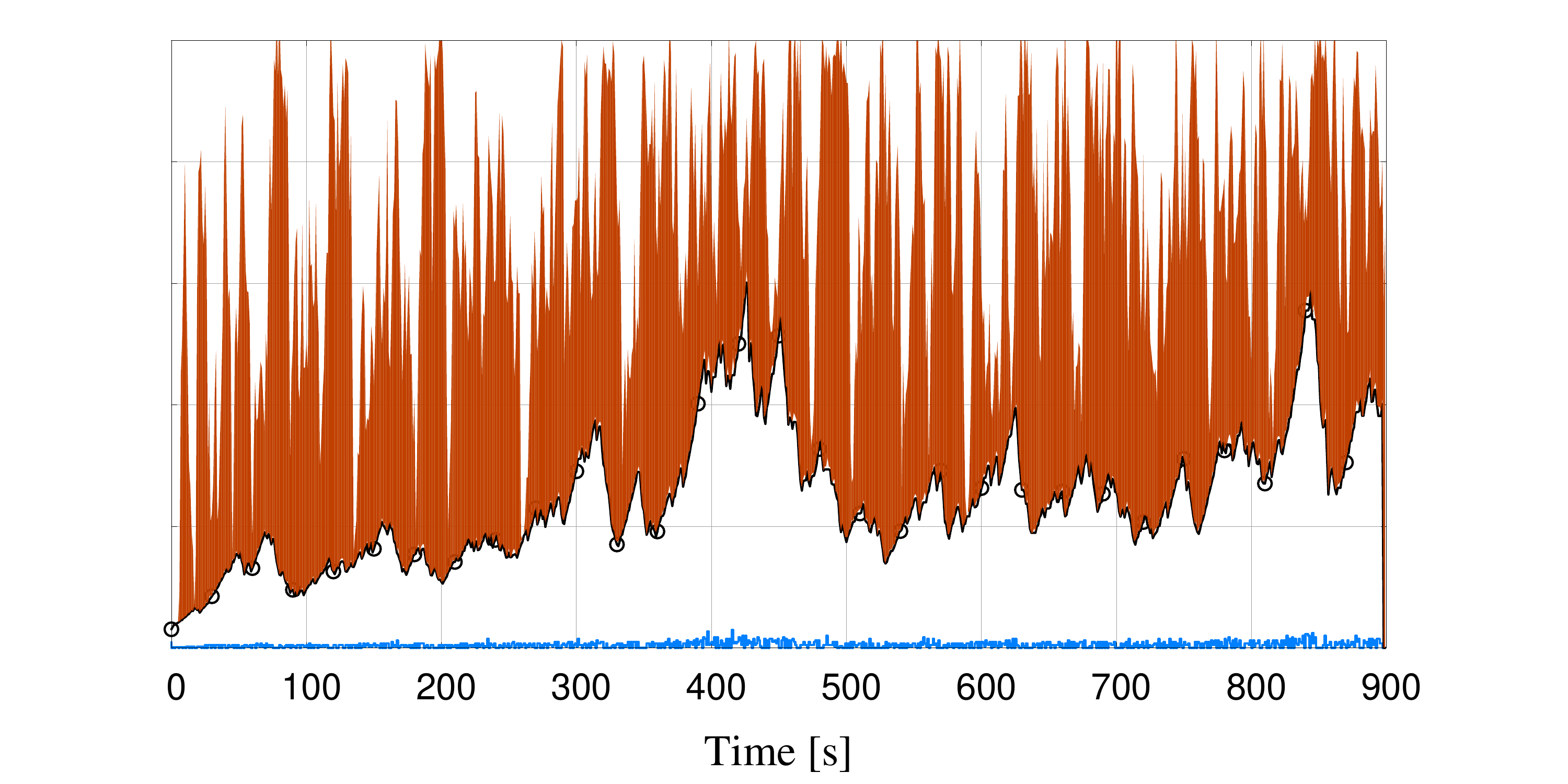}}
	\subfloat[OWD=240ms, FRCC=83.6\%]{\label{competition-time-240ms}%
	\includegraphics[scale=0.2]{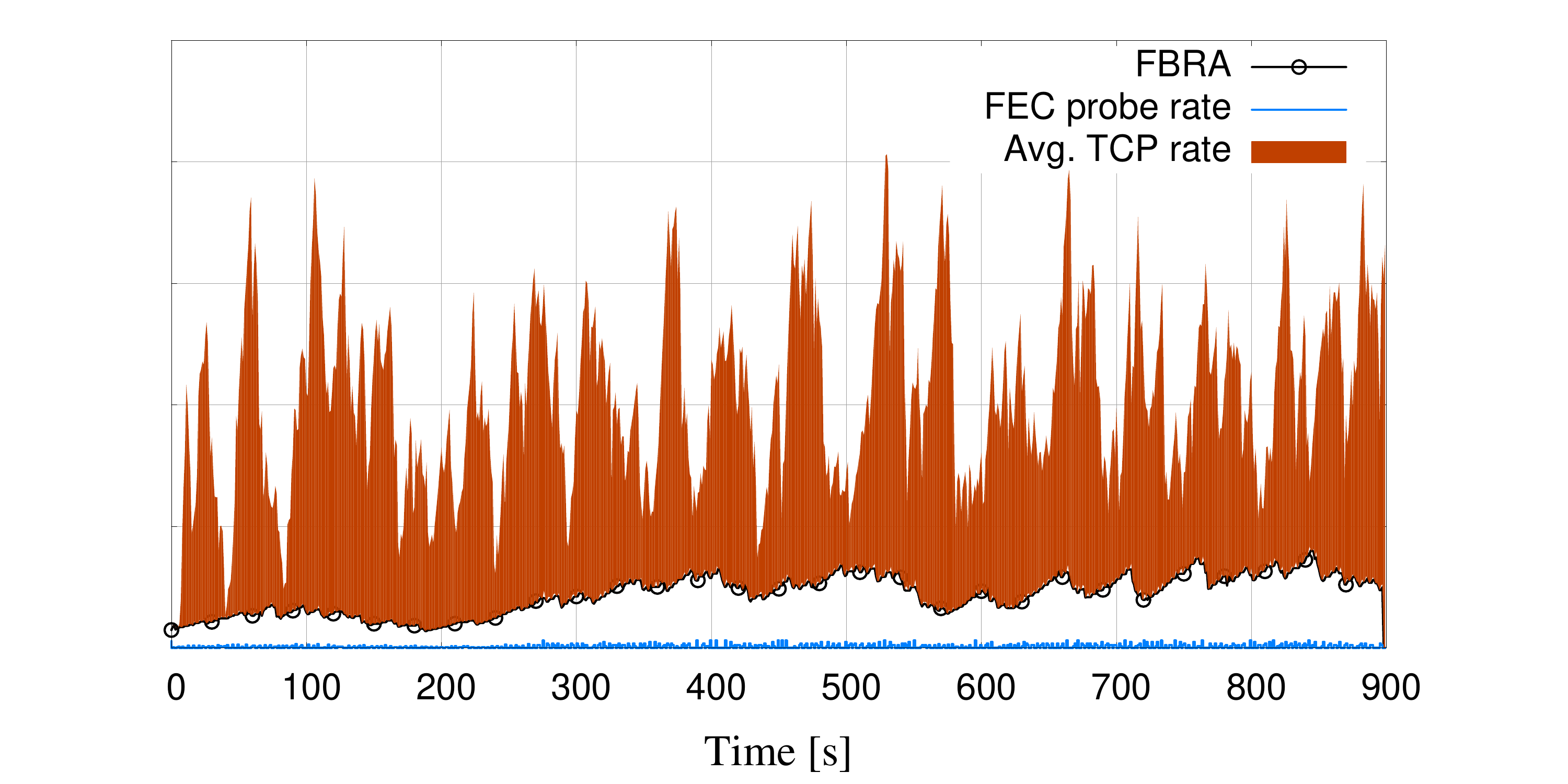}}
	\caption{The plot shows the performance of a single RTP flow using FBRA when competing 
	with 10 short TCP flows on a common bottleneck link. 
	The bottleneck link capacity is a constant $5Mbps$. To show the bottleneck 
	link utilization, the FBRA rate and the Average TCP rate are stacked on top of each other. 
	The FEC probing rate is plot independently to show that the FEC probing rate correlates with 
	the FBRA sending rate. As before, the link utilization in the very high delay ($OWD=240ms$) scenario 
	is low because 	the FBRA sense that it is operating very close to the $delay_{max}$.}
	\label{competition-graphs-1}
\end{figure*}

\subsection{Single RTP flow on a Variable Capacity Link}

Figure~\ref{scenario-1} illustrates this scenario topology, an RTP node is
simulating a video conversation with another node connected through a
constrained bottleneck link (\emph{dumbbell} topology). The access links have
a capacity of $100 Mbps$ and $1ms$ delay, while the bottleneck link capacity
varies between $100kbps$ and $256kbps$. In the scenario, we evaluate the
reactivity and convergence of the congestion control algorithm to the
available end-to-end capacity.

%the capacity at some instances changes rapidly,
%while at other instances changes rather smoothly. This is done
%to assess 

%In this simulation scenario the dumbbell topology with two RTP nodes
%simulating video conversation was used. These nodes were connected to the
%bottleneck link whose capacity was made variable. Access links had 100Mb/s
%bandwidth available, and their delay was 1ms. The reverse path was
%symmetrical.  

%The bottleneck link capacity varied between 100kbps and 256kbps and followed
%the periodic pattern presented in the figure \ref{capacity-changes}. It
%included many types of capacity changes from very rapid ones to smooth ones.
%We believe that such diverse pattern of capacity changes well verifies
%performance of presented rate control algorithms in different situations.

Our simulations show that RRTCC achieves the best goodput ($169$-$180 kbps$)
for the three bottleneck link delays (see Table~\ref{vbr-table}), but has the
worst loss rate ($3$-$4\%$). RRTCC is aggressive in its probing for available
bandwidth and as per the algorithm defined in~\cite{googleRC} does not react
to losses up to $2\%$. On the other hand, C-NADU achieves excellent
reliability ($0.15$-$0.5\%$) results in all the cases, but has lower goodput
(about $10$-$15kbps$ lower than RRTCC). The two algorithms trade-off
throughput for packet loss and vice-versa.

%FBRA performs very well for 50ms delay case and 100ms one. 

For the $50ms$ and $100ms$ bottleneck link delay, the goodput achieved by FBRA
is comparable to RRTCC but with comparatively lower loss rates
($\approx$$1.5\%$). Figures~\ref{capacity-changes}(a)--(b) show that the FBRA
can quite quickly bounce-back after undershooting. The figure also shows that
the FEC probing rate increases when the FBRA ramps up and the FEC rate is low
or disabled when the FBRA undershoots. However, FBRA is primarily a
delay-based control algorithm and for the $240ms$ bottleneck delay, it
observes that the packets are arriving very close to the $delay_{max}$ and is
therefore, conservative in its probing for available bandwidth (this is
observed by the low FEC rate in Figure~\ref{vbr-time-240ms}). FEC rate is
about 10\% of the media rate at about $15$-$20kbps$ and the accuracy of using
FEC for congestion control ($FRCC>90\%$) is also very high in each case.

\begin{table}[!t]
	\caption{Overall metrics for an RTP Flow on a Variable Capacity Link}
	\label{vbr-table}
	\resizebox{\columnwidth}{!}{%
	\begin{tabular}{| c | c | c | c | c | c | c | c |}
	\hline
	\multirow{2}{*}{\textbf{Delay}} & \multirow{2}{*}{\textbf{Metric}} & \multicolumn{2}{c|}{\textbf{FBRA}} & \multicolumn{2}{c|}{\textbf{RRTCC}} & \multicolumn{2}{c|}{\textbf{C-NADU}} \\ \cline{3-8}
	& & avg. & $\sigma$ & avg. & $\sigma$ & avg. & $\sigma$ \\ \hline
	\multirow{3}{*}{\begin{sideways} \textbf{50ms} \end{sideways}} & \textbf{Goodput [kbps]} & 179.13 & 2.26 &181.8 & 3.11& 165.42 & 3.87 \\ \cline{2-8}
	& \textbf{Loss rate [\%]} & 1.23 & 0.28 & 4.27 & 0.78 & 0.34 & 0.11\\ \cline{2-8}
	%& \textbf{FEC rate [kbps]} & 17.51 & 1.18 & - & -& -& -\\ \cline{2-8}
	& \textbf{No. of lost frames} & 441.43 & 82.37 & 1842& 25.4&93.67 & 29.64 \\ \hline
	%& \textbf{No. of FEC protected lost frames} & 91.37 & 36.55 & - & -& -& - \\ \cline{2-8}
	%&\textbf{FRCC[\%]}& 92.57 & 2.84 & - & -& -& - \\ \hline
	\multirow{3}{*}{\begin{sideways} \textbf{100ms} \end{sideways}} & \textbf{Goodput [kbps]} & 172.83 & 2.74 & 172.48 & 6.6 & 163.84 & 3.11\\ \cline{2-8}
	& \textbf{Loss rate [\%]} & 1.72 & 0.37 & 3.09&0.85 & 0.17& 0.09\\ \cline{2-8}
	%& \textbf{FEC rate [kbps]} & 20.05 & 1.34 & - & -& -& - \\ \cline{2-8}
	& \textbf{No. of lost frames} & 562.83 & 103.44 & 740& 42.82& 46.4& 22.94\\ \hline
	%& \textbf{No. of FEC protected lost frames} & 130.57 & 45.41 & - & -& -& - \\ \cline{2-8}
	%& \textbf{FRCC[\%]} & 95.85 & 1.95 & - & -& -& - \\ \hline
	\multirow{3}{*}{\begin{sideways} \textbf{240ms} \end{sideways}} &\textbf{Goodput [kbps]} & 144.89 & 8.35 & 169.22& 5.68 & 153.52& 6.81\\ \cline{2-8}
	& \textbf{Loss rate [\%]} & 2.82 & 0.89 & 2.98& 0.55&0.19 &0.07 \\ \cline{2-8}
	%& \textbf{FEC rate [kbps]} & 16.22 & 0.56 & - & -& -& - \\ \cline{2-8}
	& \textbf{No. of lost frames} & 789.93 & 223.55 & 705.67& 41.33&53.23 &21.41 \\ \hline
	%& \textbf{No. of FEC protected lost frames} & 178.07 & 69.92 & - & -& -& - \\ \cline{2-8}
	%& \textbf{FRCC[\%]} & 90.1 & 7.26 & - & -& -& - \\ \hline
	\end{tabular}
	}
\end{table}

%\subsection{Constant link capacity with one RTP flow and multiple TCP flows
%scenario}

%The purpose of this simulation experiment was to verify how well the FEC rate
%control algorithm performs if the RTP sender has to compete for resources
%with a number of TCP flows. Similarly to the previous scenario, the dumbbell
%topology with two RTP nodes simulating video conversation was used. These
%nodes together with 10 TCP nodes on each side were connected to the
%bottleneck link. The reverse path was symmetrical. Unlike the previous
%scenario, the capacity of the bottleneck link was constant and equalled
%5Mb/s. Access links had constant capacities of 100Mb/s and 1ms delay. The
%scenario topology is illustrated in the figure \ref{scenario-2}.

\begin{figure*}[!t]
	\subfloat[OWD=50ms,  FRCC=90.5\%]{\label{m-competition-time-50ms}%
		\includegraphics[scale=0.2]{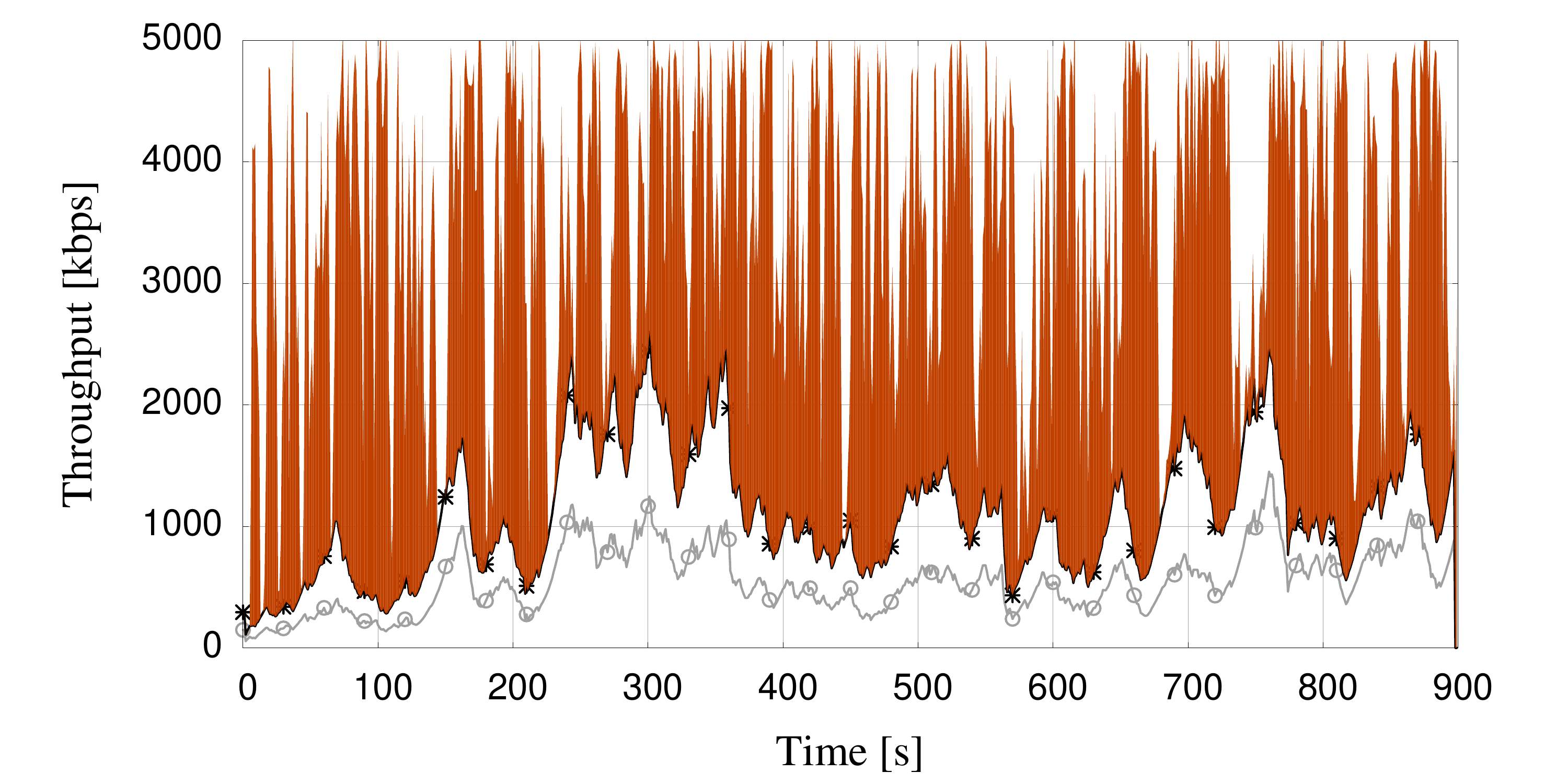}}
	\subfloat[OWD=100ms, FRCC=88\%]{\label{m-competition-time-100ms}%
		\includegraphics[scale=0.2]{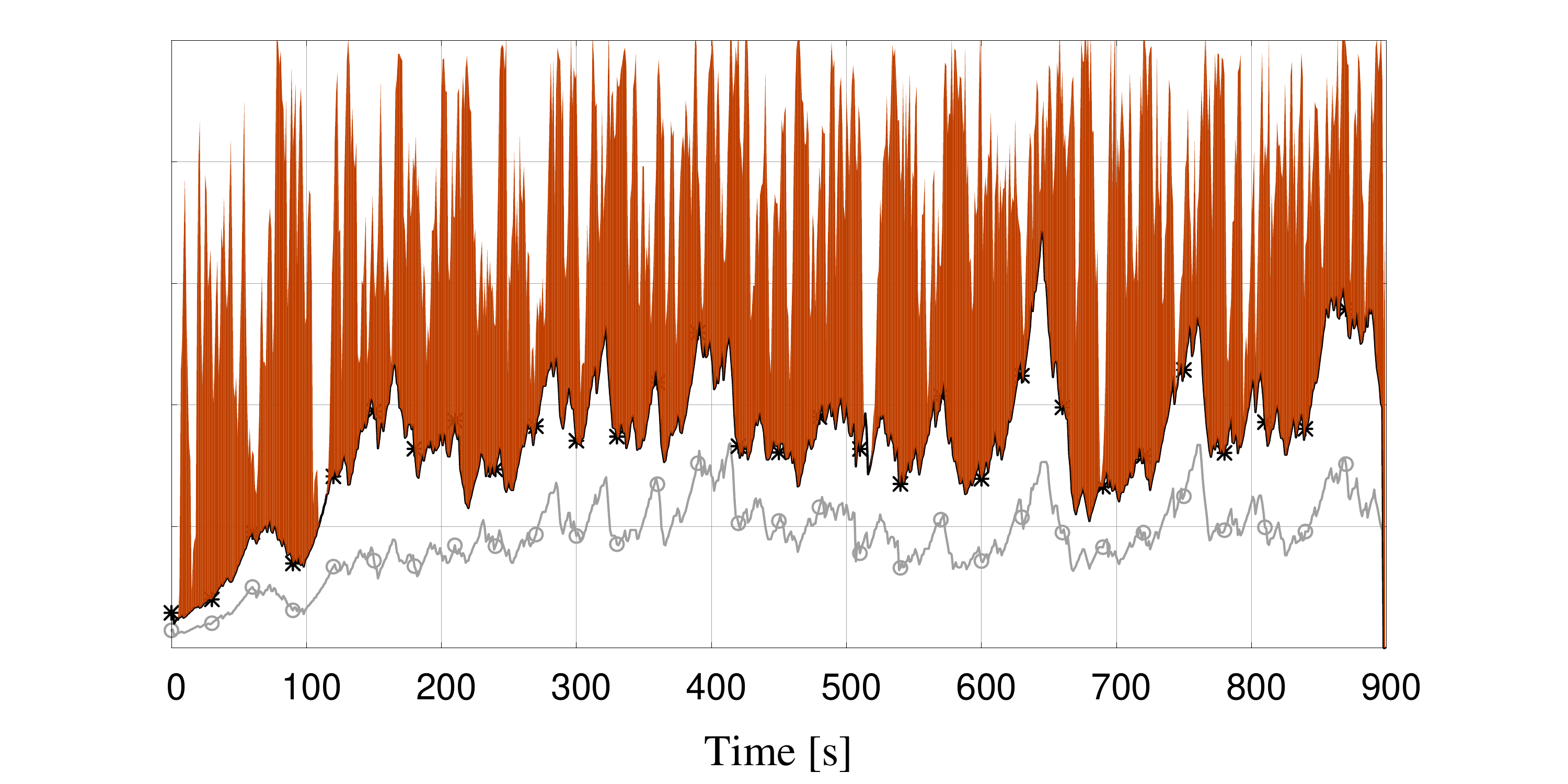}}
	\subfloat[OWD=240ms, FRCC=81.75\%]{\label{m-competition-time-240ms}%
		\includegraphics[scale=0.2]{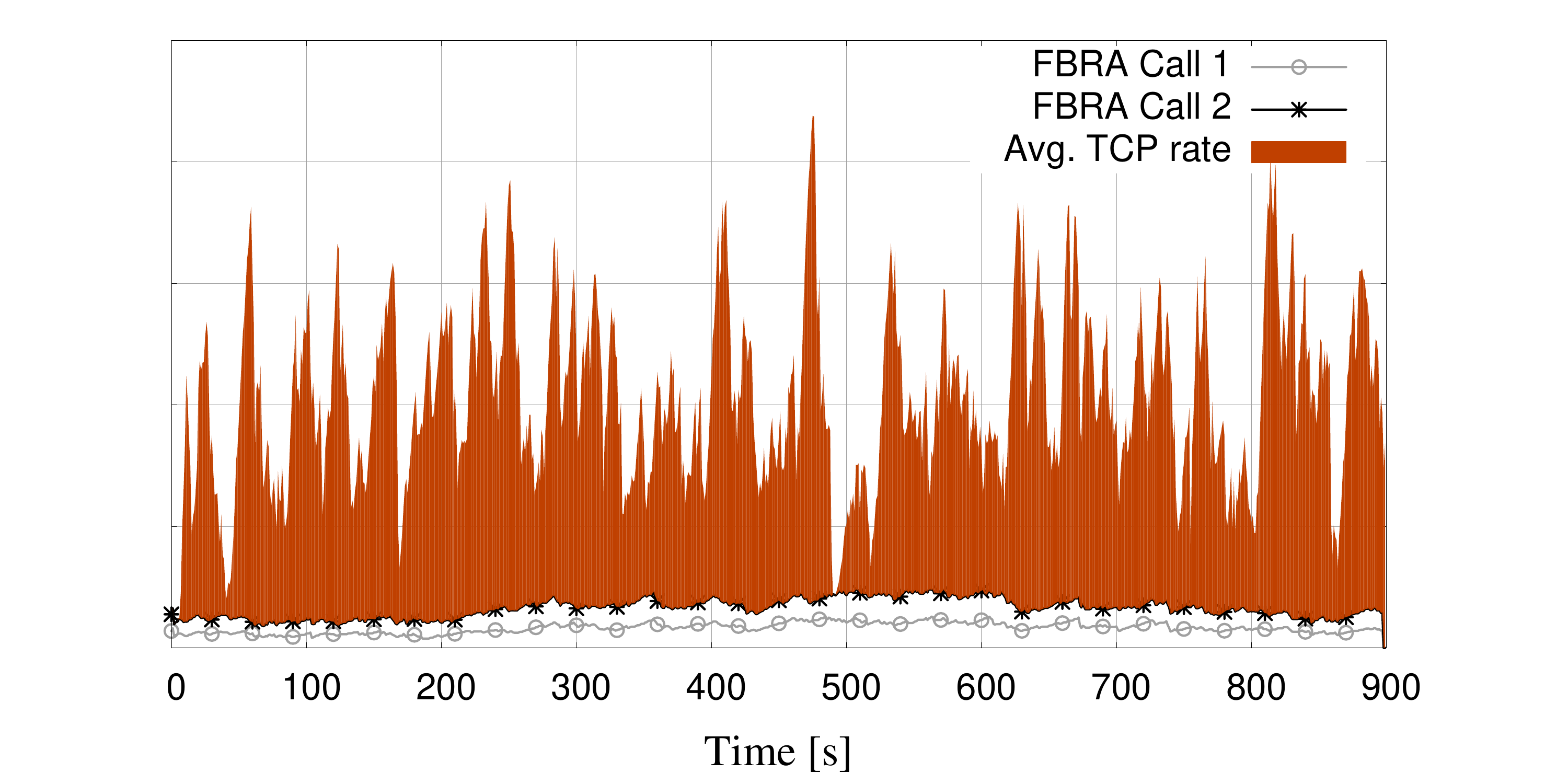}}
	\caption{The plot shows the performance of two RTP flows using FBRA competing with 
	10 short TCP flows on a bottleneck link with different delays. The sending rate for the two RTP
	flows and the average TCP throughput are stacked on top each other. The flows in all cases 
	appear fair to one another, but as observed in the previous scenarios the FBRA is 
	very conservative in probing for available bandwidth in the high delay scenario ($OWD=240ms$).}
	\label{competition-graphs-2}
\end{figure*}

\subsection{One RTP flow competing with multiple short TCP flows}
\label{1r10t}

In this simulation experiment, we evaluate the performance of the RTP flow
when it competes with a number of TCP flows. Similar to the previous scenario,
we use a dumbbell topology but instead of just a single RTP flow traversing
the bottleneck link, it is shared with $10$ short TCP connections (for e.g.,
having $10$ tabs open in a browser). Each TCP flow is modeled as a sequence of
file downloads interleaved with idle periods (on-off traffic simulating
webpage downloads). The sizes of the webpages are obtained from a uniform
distribution between $100 KB$ and $1.5MB$. Lengths of the idle periods are
drawn from an exponential distribution with the mean value of $10$ seconds.
Unlike the previous scenario, the bottleneck link has a constant available
capacity of $5Mbps$ and the topology is illustrated in the
Figure~\ref{scenario-2}.

\begin{table}[!tbp]
	\caption{Overall metrics for 1 RTP Flow competing with 10 TCP Flows}
	\label{cbr-table}
	\resizebox{\columnwidth}{!}{%
	\begin{tabular}{|c|c|c|c|c|c|c|c|}
	\hline
	\multirow{2}{*}{\textbf{Delay}} & \multirow{2}{*}{\textbf{Metric}} & \multicolumn{2}{c|}{\textbf{FBRA}} & \multicolumn{2}{c|}{\textbf{RRTCC}} & \multicolumn{2}{c|}{\textbf{C-NADU}} \\ \cline{3-8}
	& & avg. & $\sigma$ & avg. & $\sigma$ & avg. & $\sigma$\\ \hline 
	\multirow{3}{*}{\begin{sideways} \textbf{50ms} \end{sideways}} & \textbf{Goodput [kbps]} & 1044.24 & 122.9 & 3592. 9&279 & 2657.1& 164\\ \cline{2-8}
	& \textbf{Loss rate [\%]} & 1.95 & 0.26 & 3.68&0.26 &0.67 &0.15 \\ \cline{2-8}
	%& \textbf{FEC rate [kbps]} & 102.3 & 9.28& - & -& -& - \\ \cline{2-8}
	& \textbf{No. of lost frames} & 821.63 & 106.5 & 5292& 153 &1418 &134 \\ \cline{2-8}
	%& \textbf{No. of FEC protected lost frames} & 158.5 & 30.85 & - & -& -& - \\ \cline{2-8}
	%& \textbf{FFRE [\%]} & 12.63 & 3.25 & - & -& -& - \\ \cline{2-8}
	%& \textbf{FRCC[\%]} & 90.45 & 1.35 & - & -& -& - \\ \cline{2-8}
	& \textbf{TCP flow throughput [kbps]} & 669.76 & 5.41 & 688.5& 52.98& 502.28& 94.75\\ \cline{2-8}
	& \textbf{TFS [\%]} & 147.35 & 1.19 & 151.2 & 7.59& 142.27& 2.3\\ \hline
	\multirow{3}{*}{\begin{sideways} \textbf{100ms} \end{sideways}} & \textbf{Goodput [kbps]} & 1219.83 & 210.11 & 3699.43 & 419.81 & 2968.84& 59.38\\ \cline{2-8}
	& \textbf{Loss rate [\%]} & 1.54 & 0.17 & 4.05& 0.19& 0.64& 0.09\\ \cline{2-8}
	%& \textbf{FEC rate [kbps]} & 101.8 & 13.9 & - & - & - & - \\ \cline{2-8}
	& \textbf{No. of lost frames} & 685.47 & 89.34 & 6210.82&261.12 & 1422& 49.26 \\ \cline{2-8}
	%& \textbf{No. of FEC protected lost frames} & 125 & 22.06 & - & - & - & - \\ \cline{2-8}
	%& \textbf{FFRE [\%]} & 11.25 & 3.56 & - & - & - & - \\ \cline{2-8}
	%& \textbf{FRCC[\%]} & 87.76 & 1.37 & - & - & - & - \\ \cline{2-8}
	& \textbf{TCP flow throughput [kbps]} & 491.34 & 4.17 & 323.13& 110.99& 515.29 & 68.73\\ \cline{2-8}
	& \textbf{TFS [\%]} & 108.1 & 0.92 & 71.09 & 0.21 & 113.37 & 0.73\\ \hline
	\multirow{3}{*}{\begin{sideways} \textbf{240ms} \end{sideways}} & \textbf{Goodput [kbps]} & 504.82 & 82.18 & 3827.42& 423.55 & 3028.22& 78.55\\ \cline{2-8}
	& \textbf{Loss rate [\%]} & 0.4 & 0.05 & 4.82& 0.17& 0.36 &0.05 \\ \cline{2-8}
	%& \textbf{FEC rate [kbps]} & 54.12 & 7.38 & - & - & - & - \\ \cline{2-8}
	& \textbf{No. of lost frames} & 165.07 & 27.7 & 8547.93&526.8 & 597.24&123.96 \\ \cline{2-8}
	%& \textbf{No. of FEC protected lost frames} & 35.23 & 11.85 & - & - & - & - \\ \cline{2-8}
	%& \textbf{FFRE [\%]} & 18.23 & 7.94 & - & - & - & - \\ \cline{2-8}
	%& \textbf{FRCC[\%]} & 83.57 & 1.8 & - & - & - & - \\ \cline{2-8}
	& \textbf{TCP flow throughput [kbps]} & 299.12 & 2.56 &262.4 &88.34 &291.22& 37.05 \\ \cline{2-8}
	& \textbf{TFS [\%]} & 65.81 & 0.56 & 57.73 & 0.13 & 64.07 & 0.6\\ \hline
	\end{tabular}
	}
\end{table}

In our simulations, the RRTCC produces the highest goodput ($3.6$-$3.8Mbps$), 
but with high variation (see standard deviation in Table~\ref{cbr-table}). Additionally,
the RTP flow experiences high packet loss ($3.5$-$5\%$). 
%Like in the previous case,
%RRTCC is very aggressive in probing for additional capacity but in this case affects
%the throughput of the competing TCP and the TCP Fair Share (TFS) is about 
%$50$-$150\%$ (the standard deviation is also a higher at the upper bound).
%
C-NADU makes the opposite trade-off between throughput and loss rate 
and therefore, has lower goodput ($2.6$-$3Mbps$) and lower loss rate 
($0.3$-$0.7\%$). %Consequently, the TFS is a bit higher at $64$-$142\%$.

%RRTCC achieves excellent goodput results for every delay case with 3.5Mbps as
%the worst result for 50ms delay. For higher delay goodput results are better,
%but less stable, as the goodput standard deviation increases by huge margin.
%Similarly to the previous scenario, the loss rate is still significantly
%high, as it exceeds 3.6\% in every case. 

%Similarly to the previous scenario, C-NADU obtains also very good goodput
%results, and it additionally maintains very low loss rate, which never
%exceeds 0.7\%. The algorithm performs better for higher delay, as the goodput
%increases, and the results become more stable with lower goodput standard
%deviation.

Similar to the previous scenario, the FBRA algorithm performs very well in the
50ms and 100ms delay cases with goodput over $1Mbps$ and compared to RRTCC
significantly lower loss rate ($0.4$-$2.0\%$) and standard deviation of
goodput results. In the 240ms bottleneck delay case the goodput falls to
505kbps because the FBRA being delay-based becomes conservative. Furthermore,
we observe that at 240ms, FEC becomes more useful for error protection
($FFRE=18\%$) than for congestion control ($FFRC=83.6\%$). For other delays
the FFRE is lower ($11$-$12\%$) and FFRC is higher ($87$-$91\%$). In all the
scenarios the FEC rate is about $10\%$ of the media rate and we observe
recoveries due to congestion losses in this scenario, with $10$-$20\%$ of the
protected lost packets are recovered. Some packets were lost in bursts, in
these cases a parity FEC scheme cannot provide additional protection.

%Reliability results are also very good, as the worst result is obtained in
%the 50ms delay case and equals 1.95\%. Usage of FEC allows to recover around
%11\%-18\% in every case (FFRE equals 12.63\% in 50ms delay case, 11.25\% in
%100ms one, and 18.23\% in 240ms one) with FEC overhead of around 10\% in all
%cases. FBRA obtains 90.45\% of FRCC in the 50ms delay case, and it falls to
%87.76\% in the 100ms one, and further to 83.57\% in the last case. Similarly
%to the previous scenario, this result also shows that FEC applicability
%decreases for higher delays.

\subsection{Multiple RTP flows competing with multiple TCP flows}

In this simulation experiment, we add another RTP flow to the scenario
in Section~\ref{1r10t}, i.e., the second flow competes for capacity with 
the other flows at the bottleneck link. All other parameters 
including the network topology, link characteristics, and traffic source 
properties remain the same (see Figure~\ref{scenario-3} for details).

%In all three delay cases and for every algorithm, both RTP flows are fair 
%to each other, leaving the competing counterpart RTP flow enough bandwidth 
%to provide a multimedia session at a comfortable level of user experience.

\begin{table}[!t]
	\caption{Overall metrics for 2 RTP Flows competing with 10 TCP flows}
	\label{mcbr-table}
	\resizebox{\columnwidth}{!}{%
	\begin{tabular}{|c|c|c|c|c|c|c|c|c|}
	\hline
	\multirow{2}{*}{\textbf{Delay}} & & \multirow{2}{*}{\textbf{Metric}} & \multicolumn{2}{c|}{\textbf{FBRA}} & \multicolumn{2}{c|}{\textbf{RRTCC}} & \multicolumn{2}{c|}{\textbf{C-NADU}} \\ \cline{4-9}
	& & & avg. & $\sigma$ & avg. & $\sigma$ & avg. & $\sigma$\\ \hline 
	\multirow{8}{*}{\begin{sideways} \textbf{50ms} \end{sideways}} & \multirow{3}{*}{\begin{sideways} \textbf{Flow 1} \end{sideways}} & \textbf{Goodput [kbps]} & 571.21 & 94.42 & 1640.29& 37.625& 742.63&100.55\\ \cline{3-9}
	& & \textbf{Loss rate [\%]} & 2.1 & 0.21 & 5.65& 0.09& 1.2& 0.12\\ \cline{3-9}
	%& & \textbf{FEC rate [kbps]} & 62.42 & 7.98 & - &- &- &- \\ \cline{3-9}
	& & \textbf{No. of lost frames} & 746.83 & 94.84  & 7787.67 & 124.83 &801 &68.11 \\ \cline{2-9}
	%& & \textbf{No. of FEC protected lost frames} & 143.07 & 29.46  & - &- &- &- \\ \cline{3-9}
	%& & \textbf{FFRE [\%]} & 17.78 & 4.98  & - &- &- &- \\ \cline{3-9}
	%& & \textbf{FRCC[\%]} & 90.37 & 1.39  & - &- &- &- \\ \cline{2-9}
	& \multirow{3}{*}{\begin{sideways} \textbf{Flow 2} \end{sideways}} & \textbf{Goodput [kbps]} & 520.08 & 84.33 & 1750.09 & 65.45& 748.65& 72.95\\ 	\cline{3-9}
	& & \textbf{Loss rate [\%]} & 2.4 & 0.18 &5.48 &0.30 & 1.17& 0.14\\ \cline{3-9}
	%& & \textbf{FEC rate [kbps]} & 58.81 & 8.2  & - &- &- &- \\ \cline{3-9}
	& & \textbf{No. of lost frames} & 674.5 & 74.99 & 8140.33&420.82 &787.67 &128.85 \\ \cline{2-9}
	%& & \textbf{No. of FEC protected lost frames} & 136.67 & 23.61  & - &- &- &- \\ \cline{3-9}
	%& & \textbf{FFRE [\%]} & 13.84 & 3.51  & - &- &- &- \\ \cline{3-9}
	%& & \textbf{FRCC[\%]} & 90.62 & 0.94  & - &- &- &- \\ \cline{2-9}
	&  \multirow{2}{*}{\begin{sideways} \textbf{TCP} \end{sideways}} & \textbf{Throughput [kbps]} & 674.28 & 4.78 & 389.53& 36.65& 786.13& 37.80 \\ \cline{3-9}
	& & \textbf{Fair share [\%]} & 134.86 & 0.96 & 77.9 & 7.78& 152.7& 2.92 \\ \hline
	\multirow{8}{*}{\begin{sideways} \textbf{100ms delay} \end{sideways}} & \multirow{3}{*}{\begin{sideways} \textbf{Flow 1} \end{sideways}} & \textbf{Goodput [kbps]} & 745.42 & 87.85 & 1782.71& 54.49& 811.52 & 89.20\\ \cline{3-9}
	& & \textbf{Loss rate [\%]} & 1.43 & 0.2 & 5.42& 0.33 & 0.69&0.15 \\ \cline{3-9}
	%& & \textbf{FEC rate [kbps]} & 72.49 & 6.41 & - & - & - & - \\ \cline{3-9}
	& & \textbf{No. of lost frames} & 566.53 & 77.02 & 8112.67 & 480.53 & 484& 137.49\\ \cline{2-9}
	%& & \textbf{No. of FEC protected lost frames} & 94.23 & 19.53 & - & - & - & - \\ \cline{3-9}
	%& & \textbf{FFRE [\%]} & 17.16 & 5.74 & - & - & - & - \\ \cline{3-9}
	%& & \textbf{FRCC[\%]} & 87.77 & 1.2 & - & - & - & - \\ \cline{2-9}
	& \multirow{3}{*}{\begin{sideways} \textbf{Flow 2} \end{sideways}} & \textbf{Goodput [kbps]} & 691.8 & 83.58 & 1904.13&46.69& 879.89& 140.26\\ \cline{3-9}
	& & \textbf{Loss rate [\%]} & 1.67 & 0.18 & 5.3& 0.17& 0.77& 0.14\\ \cline{3-9}
	%& & \textbf{FEC rate [kbps]} & 68.79 & 6.12 & - & - & - & - \\ \cline{3-9}
	& & \textbf{No. of lost frames} & 521.57 & 68.05 & 8579.67&214.22 & 578.33& 115.8\\ \cline{2-9}
	%& & \textbf{No. of FEC protected lost frames} & 93.3 & 20.17 & - & - & - & - \\ \cline{3-9}
	%& & \textbf{FFRE [\%]} & 13.16 & 3.16 & - & - & - & - \\ \cline{3-9}
	%& & \textbf{FRCC[\%]} & 88.16 & 1.25 & - & - & - & - \\ \cline{2-9}
	& \multirow{2}{*}{\begin{sideways} \textbf{TCP} \end{sideways}} & \textbf{Throughput [kbps]} & 466.21 & 3.26 & 281.7& 23.51 & 547.63 & 32.77\\ \cline{3-9}
	& & \textbf{Fair share [\%]} & 93.24 & 0.65 & 56.2 & 7.27 & 109.4 & 3.91\\ \hline
	\multirow{8}{*}{\begin{sideways} \textbf{240ms delay} \end{sideways}} & \multirow{3}{*}{\begin{sideways} \textbf{Flow 1} \end{sideways}} & \textbf{Goodput [kbps]} & 260.26 & 100.42 & 2145.09& 18.96& 1299.87 & 364.3 \\ \cline{3-9}
	& & \textbf{Loss rate [\%]} & 0.59 & 0.19 & 6.26& 0.26& 0.47&0.19 \\ \cline{3-9}
	%& & \textbf{FEC rate [kbps]} & 30.39 & 11.78 & - & - & - & - \\ \cline{3-9}
	& & \textbf{No. of lost frames} & 155.47 & 52.85 & 11001.3 & 440.96 & 514.67& 104.71\\ \cline{2-9}
	%& & \textbf{No. of FEC protected lost frames} & 33.93 & 15.65 & - & - & - & - \\ \cline{3-9}
	%& & \textbf{FFRE [\%]} & 17.86 & 8.72 & - & - & - & - \\ \cline{3-9}
	%& & \textbf{FRCC[\%]} & 82.24 & 2.41 & - & - & - & - \\ \cline{2-9}
& \multirow{3}{*}{\begin{sideways} \textbf{Flow 2} \end{sideways}} & \textbf{Goodput [kbps]} & 287.41 & 140.73 & 2231.51 & 14.01 & 1039.08& 277.35\\ \cline{3-9}
	& & \textbf{Loss rate [\%]} & 0.71 & 0.19 & 6.02 & 0.06 & 0.54 & 0.09 \\ \cline{3-9}
	%& & \textbf{FEC rate [kbps]} & 33.67 & 15.36 & - & - & - & - \\ \cline{3-9}
	& & \textbf{No. of lost frames} & 174.37 & 61.59 & 11234.67& 120.29& 459.67& 119.68\\ \cline{2-9}
	%& & \textbf{No. of FEC protected lost frames} & 39.43 & 17.73 & - & - & - & - \\ \cline{3-9}
	%& & \textbf{FFRE [\%]} & 13.72 & 6.88 & - & - & - & - \\ \cline{3-9}
	%& & \textbf{FRCC[\%]} & 81.59 & 2.85 & - & - & - & - \\ \cline{2-9}
	& \multirow{2}{*}{\begin{sideways} \textbf{TCP} \end{sideways}} & \textbf{Throughput [kbps]} & 299.91 & 1.94 & 101.6&6.52 & 291.22& 37.05\\ \cline{3-9}
	& & \textbf{Fair share [\%]} & 59.98 & 0.39 & 20.32 & 0.10& 58.24& 0.41 \\ \hline
	\end{tabular}
	}
\end{table}

The goodput for each RTP flow is comparable to the other for all three
congestion control algorithms, i.e., each RTP flow is fair to the other and
leaves the competing RTP flow enough bandwidth to provide a similar user
experience. However, what differs between them is the interaction with the
short TCP flows. At low bottleneck link delays, the TCP Fair Share (TFS) for
TCP flows competing with FBRA and C-NADU is around or greater than $100\%$
(see Table~\ref{mcbr-table}), which shows that both algorithms move out of the
way of the TCP flows. However, RRTCC is more competitive and allows TCP flows
only $55$-$80\%$ of their fair share, which affects the completion times for
these TCP flow sand may raise some doubts about its fairness. Our observations
show that TCP flows competing with RRTCC will take $2$-$3$ times longer than
TCP flows competing with C-NADU or FBRA (compare the average TCP throughputs
in Table~\ref{mcbr-table}).

%
%As before, RRTCC provides the highest goodput but at the expense of
%packet loss. In this case the packet loss rate exceeding 5\% in every case,
%which affects the throughput of the TCP flows and the TFS is between
%$20$-$80\%$. Conversely, C-NADU has a reasonable goodput for both 
%flows and TFS
%

%well in all three scenarios with the worst goodput result of 742kbps in the
%50ms delay case, and exceeding 1Mbps in other ones. It achieves excellent
%reliability results with the highest loss rate of 1.2\% in the 50ms delay
%case.

FBRA's performance is comparable to C-NADU in the low delay scenario but due
to its delay sensing behavior performs conservatively in the high delay
scenario ($240ms$). The FEC rate is $10\%$ of the media rate and the FEC
recovers $FFRE=13$-$17$\% of the protected media packets. The FEC is more
accurate for congestion control in the low delay scenario ($FFRC=88$-$91\%$)
than for high delay ($240ms$) scenario ($FFRC=81$-$82\%$).

%reasonably well in the first two cases. Once again in the 240ms delay case,
%goodput results are poor achieving barely 260-290kbps. The loss rate is
%highest for 50ms case and equals 2.4\%, and it decreases to fall below 1\% in
%the 240ms delay case. The FEC overhead is about 10\%, as average FEC rates
%equal 58.81 and 62.42kbps, 72.49kbps and 68.79kbps, 30.39kbps and 33.67kbps
%for 50ms, 100ms and 240ms delay cases respectively. Similarly to the previous
%scenario, the FFRE maintains at the level of 13\%-17\%. Once again FRCC
%results are the best for the 50ms delay case (90.37\% and 90.62\%), and falls
%to 87.77\% and 88.16\% for the 100ms one, and further to 82.24\% and 81.59\%
%in the last one.  

%%%%%%%%%%%%%%%%%%%%%%%%%
\subsection{Real World}
\label{sec.real-world}

%- challenges with real-world implementation

%- Inaccuracy in measuring RTT using RTCP thus affecting any delay-based algorithm

%- New Design conditions (GOP, FPS)

%- Setup and the results 
\emph{Ns-2} simulations only take the network aspects into account, 
such as packet loss due to router drop, bit-error loss, queuing delay etc., 
ignoring multimedia aspects like, the Group of Picture (GOP) structure, 
different types of video frames (I- P-, or B-frames), etc. Since no real 
media packets are sent in \emph{ns-2}, the simulations can neither 
measure the PSNR nor determine the Mean-opinion Score (MoS) 
for evaluating the multimedia quality experienced by the user. 
Simulations in \emph{ns-2} also ignore the performance issues with 
real-life systems, like OS kernel, device drivers, etc. 
Therefore, we also conduct experiments in a real-world setting.

Our video call application is built on top of open-source libraries:
Gstreamer\footnote{http://gstreamer.freedesktop.org/},
x264\footnote{http://www.videolan.org/developers/x264.html} and
JRTPLib\footnote{http://research.edm. uhasselt.be/\textasciitilde jori/}. We
have also extended the JRTPLib to generate and decode FEC, perform congestion
control, generate RTCP XRs, and be compliant with RFC4585 RTCP timing rules.
In addition, the FEC module is fully compatible with RFC 5109~\cite{rfc5109}.
The application can encode and decode files or take input from a webcam and
render on the screen. To evaluate performance, we use the \emph{``News''}
video sequence\footnote{http://xiph.org} in VGA frame size and $15$ FPS.
Details concerning the system design are presented in
section~\ref{sec.system}.

Due to heterogeneity in the networks, interactive multimedia applications use
a short GOP structure ($\le
5$)~\cite{4397059,s4.eval.fw,nadu.1070341,VS:Rate}, helping overcoming
variability in the available end-to-end capacity, bit-error losses, etc. By
using FEC for congestion control, our experiments show that the congestion
control algorithms can use a very long GOP structure. Since FBRA's maximum
$FEC_{interval}$ is 14 packets, we propose using a GOP of a similar size. In
addition, the encoding/decoding complexity of a stream using a larger GOP is
much smaller than of a stream using a smaller GOP and for the same encoding
rate P-frames can be less compressed. Hence, a larger GOP stream should lead
to a better PSNR and improved energy consumption~\cite{4317642}.

%is too simple to draw any advanced conclusions, as the simulations do not include any GOP concept, and thus any user-experience metrics like PSNR cannot be calculated. Furthermore, the ns-2 simulator does not take into account some factors (e.g. OS kernel, device drivers, packet processing in intermediate nodes, etc.) which have important influence on the user experience. Therefore, we have conducted the second part of the evaluation using the Adaptive Multimedia System Toolset (AMuSys), which has been developed, and the Amazon cloud. 

%The \textbf{Adaptive Multimedia System Toolset (AMuSys)} uses open-source libraries to provide a generic framework for testing video applications. It can be divided into two main parts: multimedia and networking. The multimedia tools are implemented using open-source GStreamer library \cite{gstreamer}. The networking component has been created on top of the open-source JRTP library \cite{jrtp} developed at the Hasselt and Maastricht University. We have extended the basic library with additional modules required for FEC and the rate adaptation. The most important addition is the FEC module which is fully compliant with RFC 5109 \cite{rfc5109}. Furthermore, the library also provides extended feedback profile RFC 4585 \cite{rfc4585}, but only for point-to-point sessions. In addition, we have also added 4 extensions to RTCP headers from RFC 3611 \cite{rfc3611} (Loss RLE, Discard RLE, LRRT,DLRR). 

We use Dummynet\footnote{http://info.iet.unipi.it/
luigi/dummynet/}~\cite{Carbone:2010p3502} to emulate the variation in link
capacity, latency and/or losses in our testbed. Using Dummynet, we evaluate
the FBRA algorithm in two scenarios: 1) two RTP flows compete on a bottleneck
link, 2) RTP flow competes against short TCP flows on a bottleneck link. In
both scenarios, the bottleneck link capacity is $1Mbps$ and the end-to-end
path latencies are $50ms$ and $100ms$. The TCP traffic model is identical to
the one implemented in the simulations. Finally, we initiate a video call
between a Linux machine at Aalto University (Helsinki) and an Amazon EC2
virtual machine over the public Internet. For deriving statistical
significance, each scenario is run 10 times.

To compare our results with RRTCC in the real-world, we use Google's Chrome
browsers and the video source uses the same test YUV file instead of a webcam.
To evaluate the performance, the browsers send the media streams through our
\emph{dummynet} testbed.

%Every scenario has been run 10 times to obtain proper statistical results. In the second scenario, we stream video file over the Internet between the Amazon EC2 virtual machine in Ireland, and the Linux machine located at the Aalto University.

\begin{table}[!tbp]
	\caption{Dummynet: Two RTP flows on a bottleneck link}
	\label{realworld-udp-table}
	\resizebox{\columnwidth}{!}{%
	\begin{tabular}{|c|c|c|c|c|c|}
	\hline
	& \multirow{2}{*}{\textbf{Metric}} & \multicolumn{2}{c|}{\textbf{Call 1}} & \multicolumn{2}{c|}{\textbf{Call 2}} \\ \cline{3-6}
	& & avg. & $\sigma$ & avg. & $\sigma$ \\ \hline 
	\multirow{7}{*}{\begin{sideways} \textbf{50ms delay} \end{sideways}} & \textbf{Goodput [kb/s]} & 375.39 & 88.25 & 348.77 &  83.64 \\ \cline{2-6}
	& \textbf{Loss rate [\%]} & 1.21 & 0.19 & 1.39 & 0.69 \\ \cline{2-6}
	& \textbf{FEC rate [kb/s]} & 12.24 & 1.64 & 11.78 & 1.39 \\ \cline{2-6}
	& \textbf{No. of lost frames} & 72.5 & 7.8 & 80.1 & 27.85 \\ \cline{2-6}
	& \textbf{No. of FEC protected lost frames} & 15.3 & 3.44 & 14.6 & 2.73 \\ \cline{2-6}
	%& \textbf{No. of recovered frames} & 7.4 & 2.83 & 6.9 & 3.33 \\ \cline{2-6}
	& \textbf{FFRE [\%]} & 7.41 & 9.76 & 2.2 & 3.58 \\ \cline{2-6}
	& \textbf{FRCC [\%]} & 83.19 & 2.55 & 83.39 & 2.62 \\ \cline{2-6}
	& \textbf{PSNR [dB]} & 38.08 & 2.1 & 37.7 & 1.53 \\ \hline
	%& \textbf{Lossless PSNR [dB]} &  &  &  &  \\ \hline 
	\multirow{7}{*}{\begin{sideways} \textbf{100ms delay} \end{sideways}} & \textbf{Goodput [kb/s]} & 295.33 & 48.27 & 351.1 & 63.4 \\ \cline{2-6}
	& \textbf{Loss rate [\%]} & 3.15 & 0.93 & 2.33 & 0.87 \\ \cline{2-6}
	& \textbf{FEC rate [kb/s]} & 10.7 & 0.68 & 11.69 & 1.52 \\ \cline{2-6}
	& \textbf{No. of lost frames} & 174.6 & 92.44 & 133.8 & 42.33 \\ \cline{2-6}
	& \textbf{No. of FEC protected lost frames} & 3.0 & 2.1 & 4.1 & 3.14 \\ \cline{2-6}
	%& \textbf{No. of recovered frames} & 7.3 & 3.03 & 4.5 & 3.07 \\ \cline{2-6}
	& \textbf{FFRE [\%]} & 0.0 & 0.0 & 1.54 & 4.62 \\ \cline{2-6}
	& \textbf{FRCC [\%]} & 82.99 & 2.16 & 84.69 & 3.79 \\ \cline{2-6}
	& \textbf{PSNR [dB]} & 35.64 & 1.17 & 37.32 & 1.65 \\ \hline
	%& \textbf{Lossless PSNR [dB]} & & & & \\ \hline 
	\end{tabular}
	}
\end{table}

\textbf{Dummynet two RTP flows competition scenario}:
The RTP flows using FBRA are fair to one another (see
Table~\ref{realworld-udp-table}) and the goodput results are similar in
magnitude when compared to the simulation results. The loss rate is much lower
than in the simulation results. Hence, the frame recoveries (FFRE) are also
lower. The difference in the goodput of the calls in the two delay cases is
about $30$-$50$ \emph{kbps}, but the PSNR of these calls are similar (see
Table~\ref{realworld-udp-table}). Therefore, we conclude that small rate
variations ($\approx$ 10\%-20\%) have little bearing on the quality of the
call. The FEC rate is about $10$-$12kbps$ and is smaller than the rate
achieved in the simulations, which means that actual overhead introduced by
FEC addition is also smaller. This also implies that the rate control
algorithm remains longer in the \emph{STAY} state, thus avoiding abrupt
changes to the encoding rate, which is detrimental for user experience
\cite{Zink03subjectiveimpression}. On the other hand, RTT variations in the
physical networks causes the accuracy of using FEC for rate control (FRCC) to
be also lower than in the simulations, and is around $83$-$84\%$.

RRTCC calls have a throughput of $392$ \emph{kbps} ($\sigma=120$) and $545$
\emph{kbps} ($\sigma=130$) for $50$ \emph{ms} delay scenario and approximately
$10$-$25$ \emph{kbps} lower for the $100$ \emph{ms} delay scenario. These
results are comparable to the goodput achieved by FBRA. However, the higher
standard deviation in the bandwidth measurement of the RRTCC calls denotes
higher variation during the session and hence, poor user experience.

%The FEC rate is also much higher compared to the simulations because
%FBRA assumes that the goodput of the RTP flow is low and is more incline 
%to ramp-up quickly, but the high proportion of the FEC rate does not 
%affect the accuracy of using FEC for rate-control ($FRCC=92\%$).

\begin{figure}[!t]
\centering
	\subfloat[OWD=50ms]{\label{dummynet-50ms}\includegraphics[width=\columnwidth]{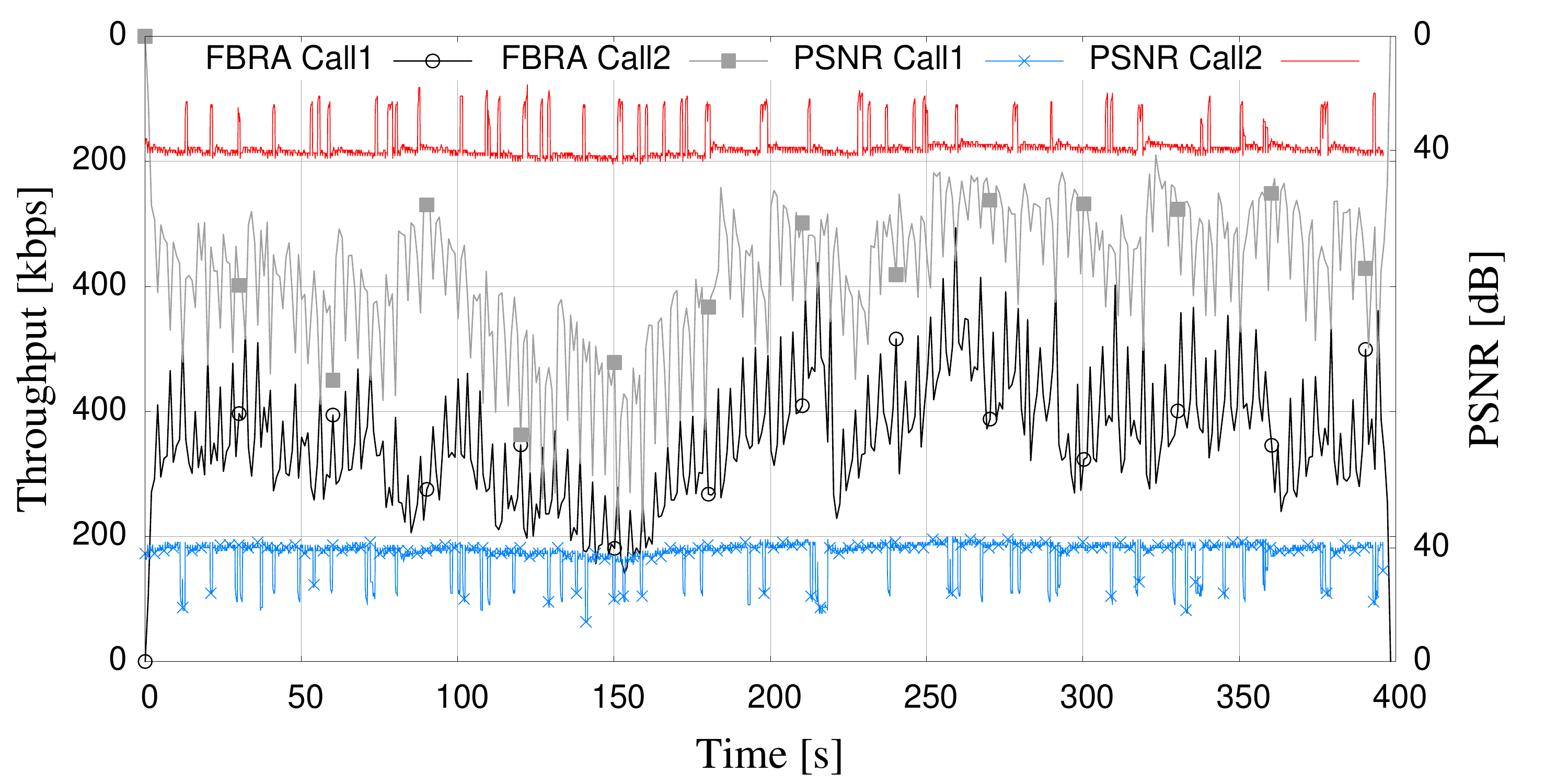}}\\
	\subfloat[OWD=100ms]{\label{dummynet-100ms}\includegraphics[width=\columnwidth]{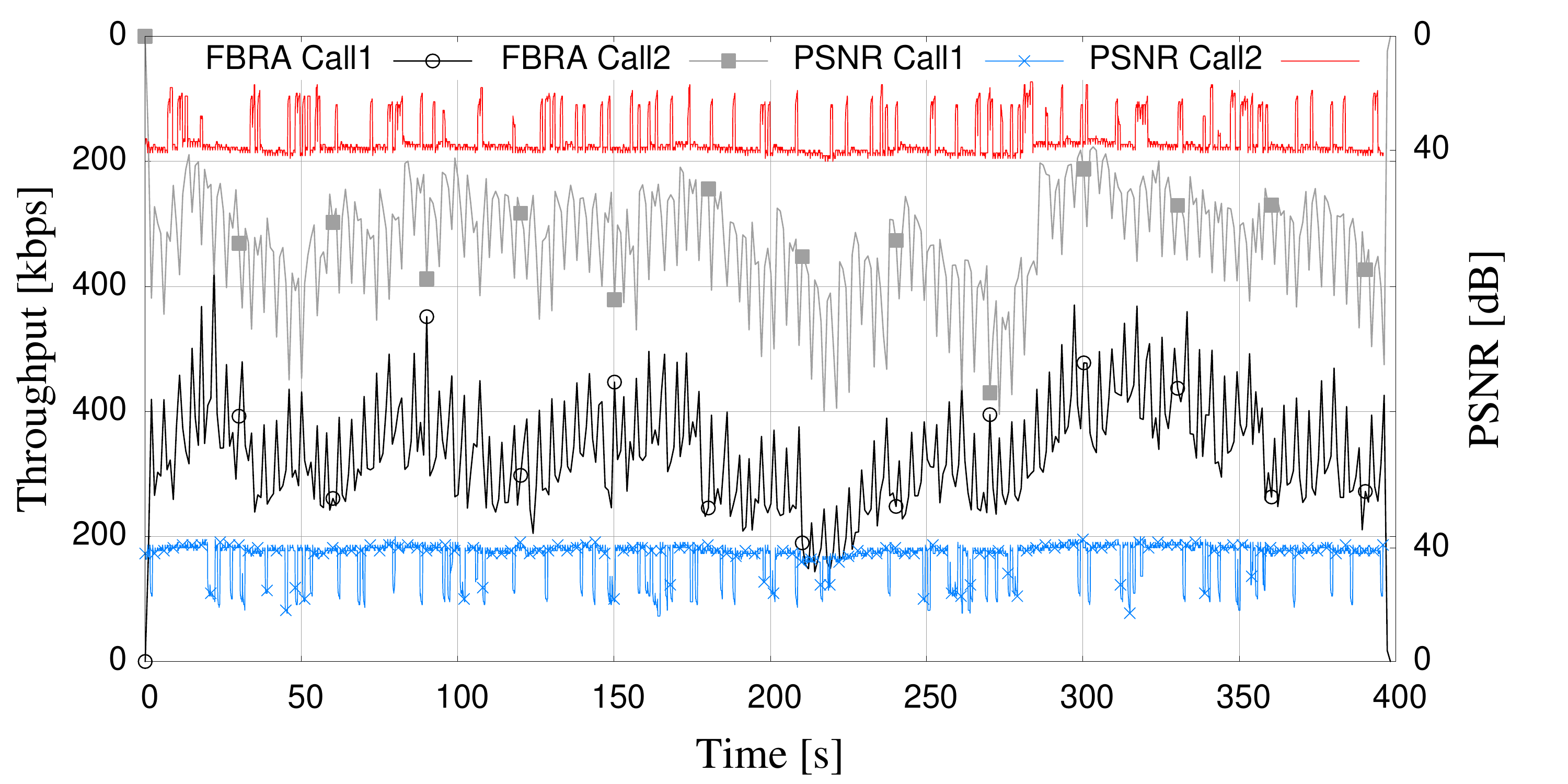}}
	\caption{Shows the goodput of two RTP calls sharing a common bottleneck. To illustrate amount of empty link capacity and how two flows push one another, we plot one of them on the reverse axis. The end-to-end path capacity is $1Mbps$ in both delay scenarios and delays are $50ms$ and $100ms$. The plot also shows the PSNR variation for the two calls (on the minor Y-axis).}
	\label{realworld-graphs}
\end{figure}

%The FBRA algorithm achieves far worse goodput results in the Dummynet in comparison to the ones obtained in the ns-2 simulator. This difference is caused by the Dummynet queuing issue described above. The goodput results improve for 100ms delay case, as they approach 200kbps for both flows. Both RTP streams are fair to each other, as they achieve comparable average goodput results for both sessions. Reliability results are much better than in the simulations, as it equals 0.06-0.07\% in the 50ms delay case, and 0.43-0.44\% in the second case. In both emulation cases PSNR results are comparable and gain around 30dB. FFRE metrics are much better than ones obtained in the simulations, but this percentage improvement results partially from far less number of lost frames. Efficiency of FEC rate control represented in the FRCC metrics is slightly better than in the simulations, as it equals around 93\% in every case. Conversely to the simulation results, FEC overhead is much higher, but this is the result of lower goodput and algorithms desire to quickly ramp-up. Scenario results are presented in the table \ref{realworld-table} and in the figure \ref{dummynet-plot}.

\begin{table}[!tbp]
	\caption{Dummynet: An RTP flow sharing a bottleneck link with short TCP flows}
	\label{realworld-tcp-table}
	\resizebox{\columnwidth}{!}{%
	\begin{tabular}{|c|c|c|c|c|}
	\hline
	\textbf{Metric} & \multicolumn{2}{c|}{\textbf{50ms delay}} & \multicolumn{2}{c|}{\textbf{100ms delay}} \\ \cline{2-5}
	& avg. & $\sigma$ & avg. & $\sigma$ \\ \hline 
	\textbf{Goodput [kb/s]} & 302.24 & 87.07 & 280.97 & 92.15 \\ \hline
	\textbf{Loss rate [\%]} & 4.24 & 0.89 & 4.1 & 0.58 \\ \hline
	\textbf{FEC rate [kb/s]} & 13.6 & 2.15 & 12.58 & 2.18 \\ \hline
	\textbf{No. of lost frames} & 154.6 & 16.56 & 170.9 & 12.38 \\ \hline
	\textbf{No. of FEC protected lost frames} & 38.0 & 6.08 & 23.7 & 8.99 \\ \hline
	% \textbf{No. of recovered frames} & 7.4 & 2.83 & 6.9 & 3.33 \\ \cline{2-6}
	\textbf{FFRE [\%]} & 6.62 & 4.01 & 6.77 & 5.8 \\ \hline
	\textbf{FRCC [\%]} & 83.32 & 2.7 & 84.06 & 2.81 \\ \hline
	\textbf{PSNR [dB]} & 35.62 & 1.49 & 34.7 & 2.26 \\ \hline
	\textbf{TCP throughput [kbit/s]} & 612.22 & 48.45 & 575.11 & 45.67 \\ \hline
	%\textbf{Lossless PSNR [dB]} &  &  &  &  \\ \hline 
	\end{tabular}
	}
\end{table}

\textbf{Dummynet RTP vs. TCP flows competition scenario}: 
The RTP flow competes well against short TCP flows achieving an average
goodput of $302$ \emph{kbps} and $280$ \emph{kbps} in the $50ms$ and $100ms$
delay scenario, respectively (see Table~\ref{realworld-tcp-table}). The TCP
flow achieves a throughput of around $600$ \emph{kbps} on average, the loss
rate is around $4\%$, which is higher than in the simulation results. This
effect mainly arises from the variations in RTT, which leads to a higher
amount of packet discards at the receiver. Frame recoveries are also less
frequent with \emph{FFRE} $\approx 6$-$7\%$.
PSNR results obtained in both scenarios are very similar ($\approx 35dB$), the
PSNR is lower for the $100ms$ delay scenario because the average goodput is
also a bit lower in this case.
Similar to the previous scenario, the FEC rate is about $12$-$13kbps$ and the
\emph{FRCC}$\approx 83$-$84\%$. This is lower than in the ns-2 simulations,
and is again due to the RTT variations, which are not present in the
simulation environment.

RRTCC calls have a throughput of $203$ \emph{kbps} ($\sigma=26$) in the $50$
\emph{ms} delay scenario, with TCP throughput of $761$ \emph{kbps}
($\sigma=238$). In the $100$ \emph{ms} case, RRTCC obtains $189$ \emph{kbps}
($\sigma=23$), while TCP reaches $867$ \emph{kbps} ($\sigma=236$).

\begin{figure}[!ht]
\centering
	\subfloat[Two RTP competition scenario]{\label{rrtcc-delay-1}\includegraphics[width=\columnwidth]{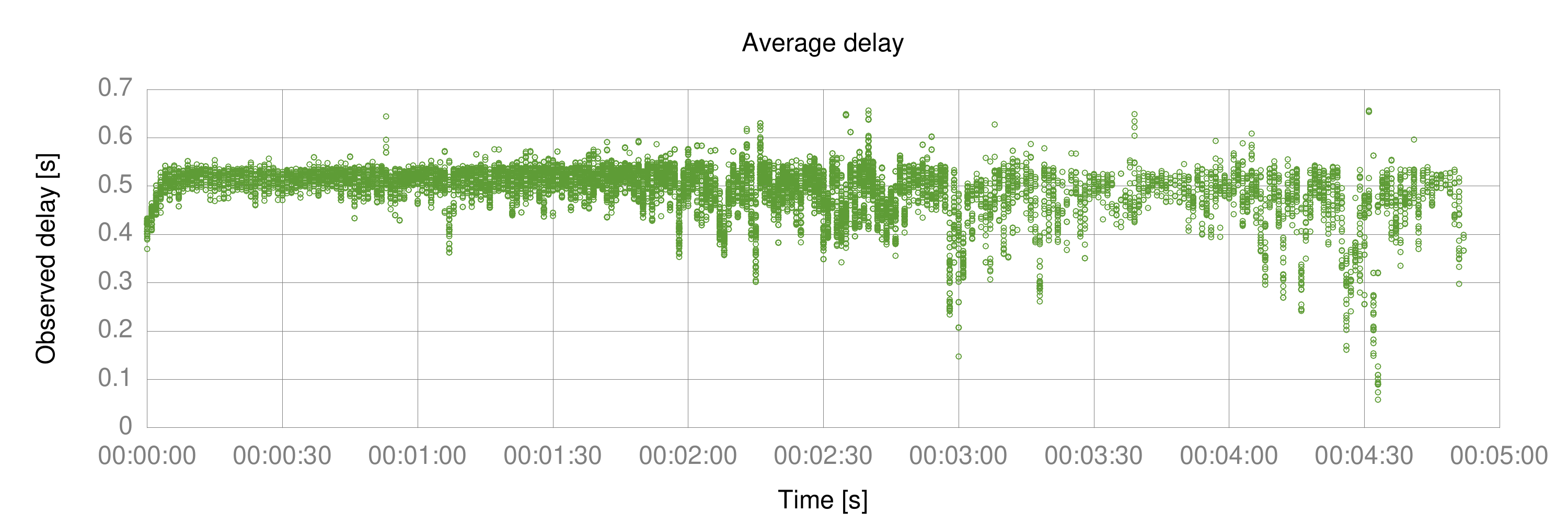}}\\
	\subfloat[RTP vs. TCP competition scenario]{\label{rrtcc-delay-2}\includegraphics[width=\columnwidth]{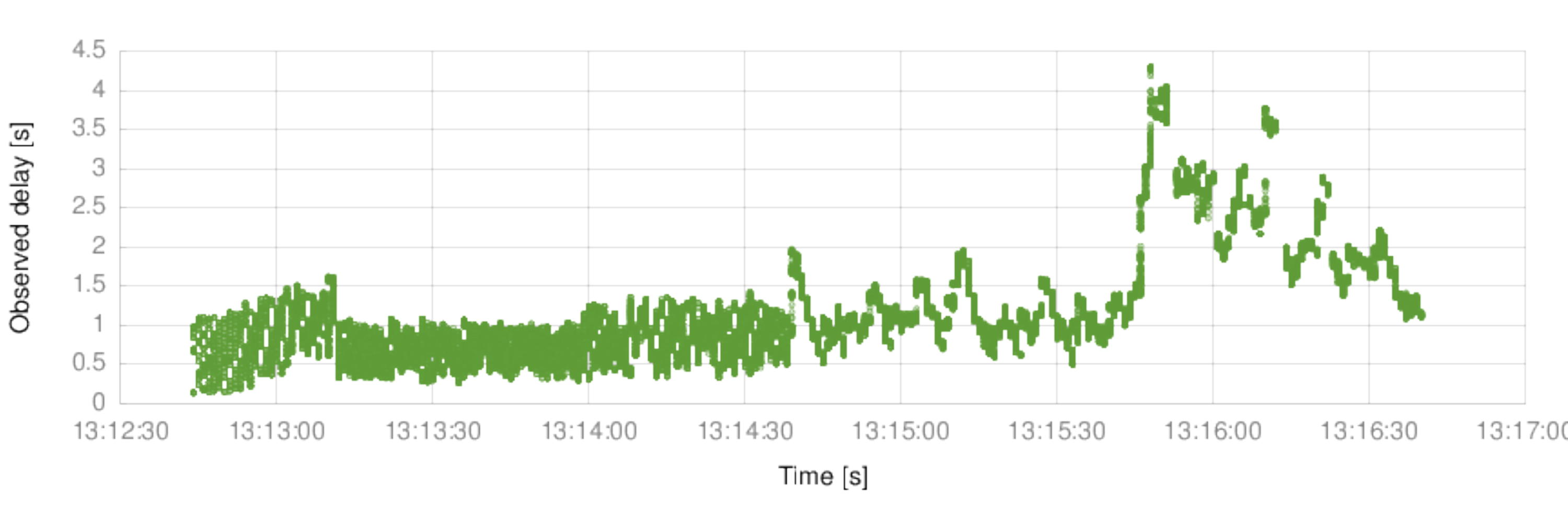}}
  \caption{shows the variation in packet delay for RRTCC congestion control.
  We observe packet delay much higher than the \emph{100ms} link latency.
  Furthermore, the observed end-to-end delay rarely goes below the recommended
  \emph{400ms}, and sometimes the delay spikes to \emph{3s}, which makes video
  conversation effectively impossible. This is a representative plot of 10
  successive runs using the Chrome browser on our testbed.}
	\label{rrtcc-delay-graphs}
\end{figure}

Higher standard deviation is not the only metric in which FBRA outperforms
RRTCC. While RRTCC flows achieve high throughput, they also induce higher
packet delay. Figure~\ref{rrtcc-delay-graphs} illustrates the variation in the
observed packet delay: 1) when two RRTCC flows compete with one another, and
2) when an RRTCC flow competes with a short TCP flow on a bottleneck link. We
observe that the packet delay constantly exceeds the recommended \emph{400ms}
end-to-end delay~\cite{s4.eval.fw} and sometimes spikes to values as high as
\emph{3s} (see figure~\ref{rrtcc-delay-2}). On the other hand, FBRA always
maintains packet delay below \emph{400ms}, and discards all packets that
arrive later. As a result, despite very comparable results in terms of
throughput, FBRA provides better user experience, as the throughput variations
are smoother, and the packet delay variation is lower.

\textbf{Call over the public Internet}: 
By initiating a video call between a host on an Amazon EC2 instance and a
machine at the university, we measure the performance of the FBRA in the
public Internet. We observe varying results between each successive run, as
the public Internet has varying amount of cross traffic. The goodput ranges
between $100$-$700$ \emph{kbps} and the maximum loss rate does not exceed
$1.5\%$. The PSNR of the calls also varies between $35$-$40$ \emph{dB}.
Despite results diversity, we show that FBRA may work in the public
Internet.

%In a series of experiments with the Amazon cloud, we have achieved excellent reliability results with loss rate never exceeding 1.5\%. Goodput results have varied between 100kbps and 700kbps, as they highly depend on the network traffic which is out of our control. Goodput variation influences PSNR results, as they also differ between around 25dB and 40dB. Because we have not enough knowledge about network situation during the experiments, we are unable to verify how well the algorithm utilizes available network bandwidth. The only conclusion drawn by us is the fact of correct response to congestion cues, and therefore proper congestion control functionality. 
%\begin{table}
%	\caption{Overall metrics for the Amazon results}
%	\label{amazon-table}
%	%\resizebox{\columnwidth}{!}{%
%	\begin{tabular}{|c|c|}
%	\hline
%	\textbf{Metric} & \textbf{Value} \\ \hline
%	\textbf{Goodput [kb/s]} & 707.27  \\ \hline
%	\textbf{Loss rate [\%]} & 0.06 \\ \hline
%	\textbf{FEC rate [kb/s]} & 281.95 \\ \hline
%	\textbf{No. of lost frames} & 1 \\ \hline
%	\textbf{No. of FEC protected lost frames} & 0 \\ \hline
%	%& \textbf{No. of recovered frames} & & & & \\ \cline{2-6}
%	\textbf{FFRE [\%]} & 0 \\ \hline
%	\textbf{FRCC [\%]} & 96.23 \\ \hline
%	\textbf{PSNR [dB]} & 38.97 \\ \hline
%	\end{tabular}
%	%}
%\end{table}

\section{System considerations}
\label{sec.system}

Congestion control algorithms for multimedia communication are never used
stand-alone, but are built into advanced systems where tight co-operation
between multiple components is required. In this section, we present the
design and implementation of the \emph{Adaptive Multimedia System} (AMuSys).

\subsection{System Description}

\begin{figure}[!t]
	\centering
	\resizebox{0.96\columnwidth}{!}{
	\includegraphics[width=\columnwidth, trim=2cm 2cm 0cm 0cm]{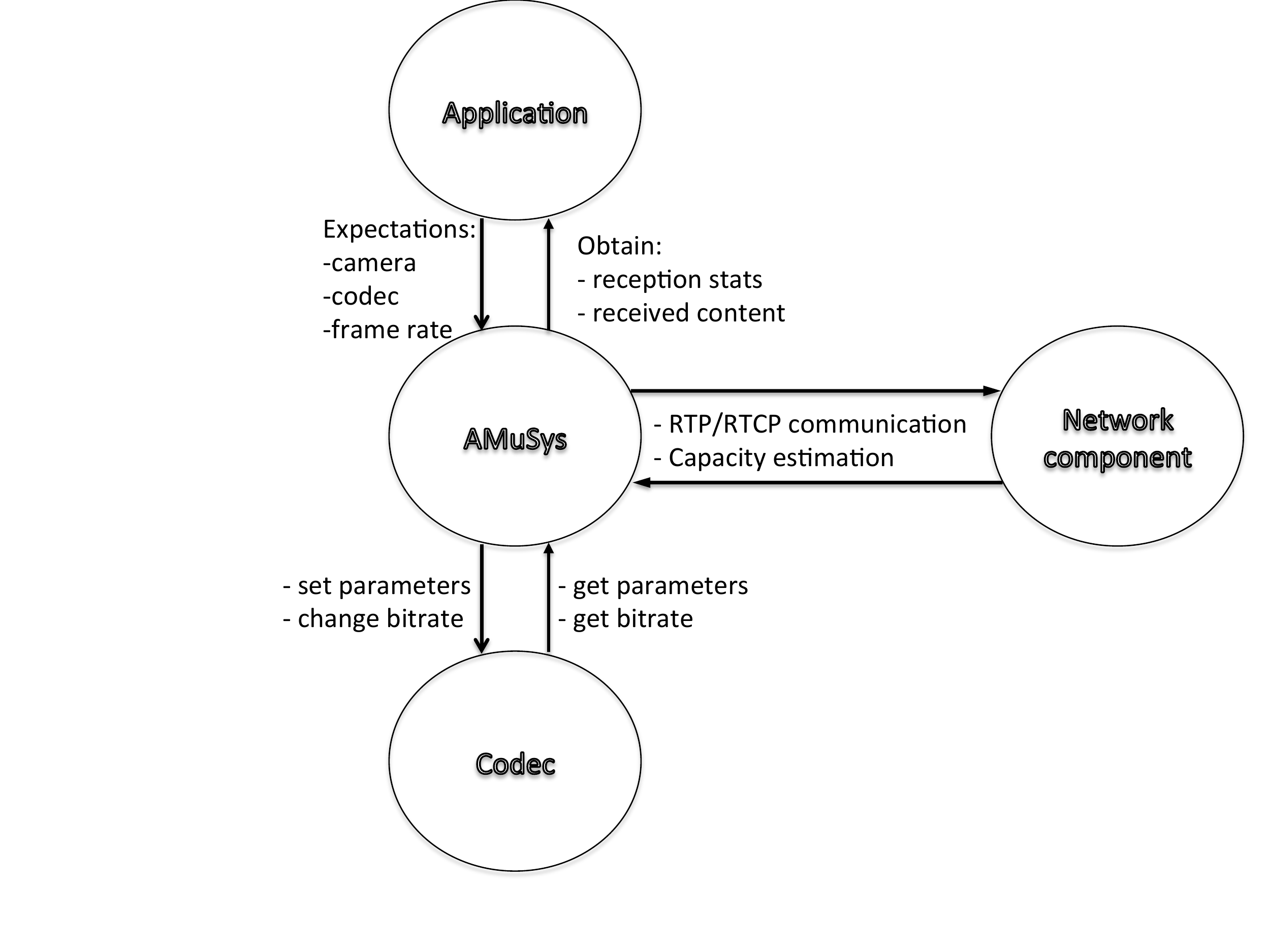}
}
    \caption{shows the design of the \emph{Adaptive Multimedia System
    (AMuSys)}. The system provides APIs to communicate between the
    application, the network component and the codec.}
	\label{amusys-descr}
\end{figure}

\emph{AMuSys} is made up of 3 sub-systems, namely, the application, codecs and
the networking components. Figure~\ref{amusys-descr} illustrates the
integration of the application, codec, and the network subsystems in
\emph{AMuSys}. It also implements the control loops (presented
in~\cite{perkins12}) needed to design advanced congestion control algorithms
that take into consideration the codec constraints along with the
network-related parameters. Application developers, who may not be multimedia
communication experts, may delegate the responsibility for the whole
communication process to \emph{AMuSys} by just specifying the desired
application preferences. \emph{AMuSys} defines three main interfaces, namely
the application interface, the network interface, and the codec interface.
Table~\ref{tab:amusys-iface} summarizes the methods offered by each interface.

\begin{table*}[!t]
\centering
\resizebox{\textwidth}{!}{%
\begin{tabular}{|l|c|c|l|}
\hline
\bf{Name} & \bf{Input} & \bf{Output} & \bf{Description} \\ \hline
\multicolumn{4}{|c|}{\emph{\textbf{Application interface}}} \\ \hline
\texttt{setPreferences()} & \texttt{dictionary} & - & Specifies end-user application requirements \\ \hline
\multicolumn{4}{|c|}{\emph{\textbf{Codec interface}}} \\ \hline
\texttt{getParam()} & \texttt{name} & \texttt{value} & Gets value of parameter name \\ \hline
\texttt{setParam()} & \texttt{name, value} & \texttt{true/false} & Sets parameter name to value \\ \hline
\texttt{getBitrate()} & - & \texttt{value} & Gets current bitrate \\ \hline
\texttt{setBitrate()} & \texttt{value} & - & Sets bitrate to $value$ \\ \hline 
\multicolumn{4}{|c|}{\emph{\textbf{Network interface}}} \\ \hline
\texttt{onReceivedRTPPacket()} & \texttt{payload, source address, timestamp} & - & Invoked on RTP packet reception\\ \hline
\texttt{onReceivedRTCPPacket()} & \texttt{payload, source address, timestamp} & - & Invoked on RTCP packet reception \\ \hline
\texttt{onSendRTPPacket()} & \texttt{payload, timestamp} & - & Invoked after RTP packet sending\\ \hline
\texttt{onSendFECPacket()} & \texttt{payload, timestamp} & - & Invoked after FEC packet sending \\ \hline
\texttt{onReceivedFECPacket()} & \texttt{payload, source address, timestamp} & - & Invoked on FEC packet reception \\ \hline
\texttt{enableFEC()} & \texttt{true/false} & - & Switches on/off FEC\\ \hline
\texttt{setFECScheme()} & \texttt{Scheme specific input} & - & Updates used FEC scheme \\ \hline
\texttt{setSessionParams()} & \texttt{dictionary} & - & Specifies session parameters \\ \hline
\texttt{sendEarlyRTCPReport()} & - & - & Sends early RTCP packet \\ \hline
\end{tabular}
}
  \caption{Methods exposed by the application, codec and network subsystems of
  \emph{AMuSys}.}
  \label{tab:amusys-iface}
\end{table*}

The application interface allows an end-user application to specify its
preferences by calling the \texttt{setPreferences()} method. The
typical preferences are:
\begin{compactenum}
\item[(1)] codec type,
\item[(2)] expected frame rate,
\item[(3)] signal source device,
\item[(4)] expected display resolution.
\end{compactenum}
These preference may also be a result of capability negotiation during session
setup between the endpoints

The codec interface allows the \emph{AMuSys} to modify the codec settings
during an ongoing multimedia session. The interface is designed to be flexible
and codec independent. Therefore, all codec parameters are set using the
\texttt{setParam()} method, which takes 2 arguments, namely a key-value pair
containing the name of the codec-parameter, and the updated value. Since the
method is codec-independent, it is the responsibility of the systems
integrator to use the correct parameter name and value. Using incorrect
arguments causes a non-fatal error to be reported and the parameter is not
modified. The interface also provides a \texttt{getParam()} method to be able
to fetch the current value of a desired parameter. In addition, the interface
explicitly provides a method to obtain and update the media encoding rate. The
\texttt{setBitrate()} method updates the encoding rate, while the
\texttt{getBitrate()} method can be used to check if the requested rate has
been actually imposed by the encoder, or to monitor how quickly the codec is able
to converge its encoding rate to the requested value.

The network interface provides functions to modify the network parameters of
the multimedia session. The features are divided into two main categories. The
first one comprises callback functions that are invoked in the \emph{AMuSys}
control unit by the network component (e.g., RTP library) when a specific
event occurs. For instance, it may process the RTCP RR in the
\texttt{onReceivedRTCPPacket()} callback. The most important callback
functions are listed in the table~\ref{tab:amusys-iface}. The second category
consists of basic session settings that can be updated during runtime. These settings
include:
\begin{compactenum}
\item[(1)] session profile choice (AVP~\cite{rfc3550} vs. AVPF~\cite{rfc4585}), 
\item[(2)] RTCP interval modifications, 
\item[(3)] RTCP extensions choice (e.g., XR blocks~\cite{rfc3611}, 
\item[(4)] rapid timestamp synchronization (RFC 6051~\cite{rfc6051}).
\end{compactenum}

Since the \emph{AMuSys} is able to access information from the other
interfaces, it has a better understanding of the current system state, i.e.,
the current network, codec and application state, and thus is able to make
better decisions for providing good quality user-experience.

%%%%%%%%%%%%%%%%%%%%%%%%%

\section{Conclusion}
\label{sec.concl}

In this paper, we show that FEC can be applied not only for error resilience,
but can also be used for congestion control. We present a new congestion
control algorithm (FBRA) that incorporates FEC packets for probing for
available capacity. Performance of FBRA is compared with two other congestion
control algorithms, namely, RRTCC and C-NADU. Our simulations show that RRTCC
and C-NADU make opposite performance trade-offs (higher capacity instead of
lower packet loss rate, and vice-versa). The FBRA algorithm evaluation shows
that it can perform at a trade-off point between the other two algorithm. FBRA
on average has better goodput than C-NADU and better packet loss rate than
RRTCC. We also note that FBRA's performance drops at high e2e delay ($240ms$),
because FBRA uses OWD to sense congestion. Also the applicability of FEC for
congestion control reduces at high link delays, which is visible in worsening
FRCC metric.

% (accuracy of FEC for rate-control)

Furthermore, we evaluate the performance of FBRA and RRTCC in the real-world
scenarios. We show that despite obtaining comparable throughput, FBRA provides
users with enhanced user experience, as its packet delay variation and goodput
variations are far lower.

Finally, we measure the performance of FBRA on the public Internet and
show that the algorithm can be successfully applied. Since using FEC increases
error resilience, we are able to increase the GOP size, which reduces
the encoding and decoding complexity without affecting the user-experience.

%As we are interested in the general applicability of FEC,

As our concept is targeted to work on top of a congestion control unit, we
believe that it can be applied not only to the FBRA algorithm, but also to any
other rate control algorithm. This idea can be realized by adding an extra FEC
subsystem to the congestion controller. For instance in RRTCC, when it
receives a TMMBR message for increasing the rate, it can allocate the
difference in the current rate and the new rate to FEC.

In the future work, we envision exploring incorporation of more complex FEC
schemes. As the FEC frame recovery metric still shows room for improvement, we
believe that application of different FEC scheme can provide useful gain to
the overall performance. Furthermore, we are also interested in applying our
concept in challenging heterogeneous environments where FEC features can prove
to be particularly useful.

%FEC can be successfully used not only as the error resilience mechanism, but also for rate control of multimedia streams. We have shown in various simulation scenarios that the FBRA algorithm can reasonably compete with other existing congestion control algorithms, especially in the multimedia sessions with delay up to 100ms. Its performance drops, as the delay increases, which is visible in deteriorating FRCC results. Furthemore, we have shown in the real world section \ref{sec.real-world} that the algorithm can be successfully deployed in the Internet with GOP size being increased three times to 12. We claim that by usage of FEC together with larger GOP, user-experience is not deteriorated, as the FBRA algorithm performs reasonably well, and stress on encoder is decreased, which leads to improved energy efficiency.




\bibliographystyle{IEEEtran}
%\bibliography{IEEEabrv,../allpapers}
\bibliography{IEEEabrv,../allpapers,../rfc}

\begin{thebibliography}{10}
\providecommand{\url}[1]{#1}
\csname url@samestyle\endcsname
\providecommand{\newblock}{\relax}
\providecommand{\bibinfo}[2]{#2}
\providecommand{\BIBentrySTDinterwordspacing}{\spaceskip=0pt\relax}
\providecommand{\BIBentryALTinterwordstretchfactor}{4}
\providecommand{\BIBentryALTinterwordspacing}{\spaceskip=\fontdimen2\font plus
\BIBentryALTinterwordstretchfactor\fontdimen3\font minus
  \fontdimen4\font\relax}
\providecommand{\BIBforeignlanguage}[2]{{%
\expandafter\ifx\csname l@#1\endcsname\relax
\typeout{** WARNING: IEEEtran.bst: No hyphenation pattern has been}%
\typeout{** loaded for the language `#1'. Using the pattern for}%
\typeout{** the default language instead.}%
\else
\language=\csname l@#1\endcsname
\fi
#2}}
\providecommand{\BIBdecl}{\relax}
\BIBdecl

\bibitem{gettys:2012:bufferbloat}
J.~Gettys and K.~Nichols, ``Bufferbloat: dark buffers in the {Internet},''
  \emph{in Proc. of Communications of the {ACM}}, vol.~55, pp. 57--65, Jan
  2012.

\bibitem{Zink03subjectiveimpression}
M.~Zink, O.~K\"{u}nzel, J.~Schmitt, and R.~Steinmetz, ``{Subjective Impression
  of Variations in Layer Encoded Videos},'' in \emph{Proc. of IWQoS}, 2003.

\bibitem{jennings:2013:webrtc}
C.~Jennings, T.~Hardie, and M.~Westerlund, ``Real-time communications for the
  web,'' \emph{{IEEE} Communications Magazine}, vol.~51, no.~4, pp. 20--26,
  April 2013.

\bibitem{draft.rtp.cb}
C.~Perkins and V.~Singh, ``{RTP Congestion Control: Circuit Breakers for
  Unicast Sessions},'' 2012, {IETF Internet Draft}.

\bibitem{pv-2013-cb}
V.~Singh, S.~McQuistin, M.~Ellis, and C.~Perkins, ``{Circuit Breakers for
  Multimedia Congestion Control},'' in \emph{Proc. of IEEE Packet Video 2013},
  2013.

\bibitem{Devadoss2008}
J.~Devadoss, V.~Singh, J.~Ott, C.~Liu, Y.-K. Wang, and I.~Curcio, ``{Evaluation
  of Error Resilience Mechanisms for 3G Conversational Video},'' in \emph{Proc.
  of IEEE ISM}, 2008, pp. 378--383.

\bibitem{664283}
Y.~Wang and Q.-F. Zhu, ``Error control and concealment for video communication:
  a review,'' \emph{in Proc. of the IEEE}, vol.~86, no.~5, pp. 974 --997, may
  1998.

\bibitem{855913}
Y.~Wang, S.~Wenger, J.~Wen, and A.~Katsaggelos, ``Error resilient video coding
  techniques,'' \emph{in Proc. of IEEE Signal Processing Magazine}, vol.~17,
  no.~4, pp. 61 --82, jul 2000.

\bibitem{5953578}
J.~Evans, A.~Begen, J.~Greengrass, and C.~Filsfils, ``Toward lossless video
  transport,'' \emph{in Proc. of IEEE Internet Computing}, vol.~15, no.~6, pp.
  48 --57, nov.-dec. 2011.

\bibitem{ns2}
``{The Network Simulator NS-2},'' \url{http://www.isi.edu/nsnam/ns/}.

\bibitem{rfc3550}
\BIBentryALTinterwordspacing
H.~Schulzrinne, S.~Casner, R.~Frederick, and V.~Jacobson, ``{RTP: A Transport
  Protocol for Real-Time Applications},'' RFC 3550 (INTERNET STANDARD),
  Internet Engineering Task Force, Jul. 2003, updated by RFCs 5506, 5761, 6051,
  6222, 7022. [Online]. Available: \url{http://www.ietf.org/rfc/rfc3550.txt}
\BIBentrySTDinterwordspacing

\bibitem{tcp-real-time}
E.~Brosh, S.~A. Baset, D.~Rubenstein, and H.~Schulzrinne, ``{The
  Delay-Friendliness of TCP},'' in \emph{Proc. of ACM SIGMETRICS}, 2008.

\bibitem{rfc4585}
\BIBentryALTinterwordspacing
J.~Ott, S.~Wenger, N.~Sato, C.~Burmeister, and J.~Rey, ``{Extended RTP Profile
  for Real-time Transport Control Protocol (RTCP)-Based Feedback (RTP/AVPF)},''
  RFC 4585 (Proposed Standard), Internet Engineering Task Force, Jul. 2006,
  updated by RFC 5506. [Online]. Available:
  \url{http://www.ietf.org/rfc/rfc4585.txt}
\BIBentrySTDinterwordspacing

\bibitem{draft.rmcat.feedback}
C.~Perkins, ``{On the Use of RTP Control Protocol (RTCP) Feedback for Unicast
  Multimedia Congestion Control},'' 2013, {IETF Internet Draft}.

\bibitem{rfc3611}
\BIBentryALTinterwordspacing
T.~Friedman, R.~Caceres, and A.~Clark, ``{RTP Control Protocol Extended Reports
  (RTCP XR)},'' RFC 3611 (Proposed Standard), Internet Engineering Task Force,
  Nov. 2003. [Online]. Available: \url{http://www.ietf.org/rfc/rfc3611.txt}
\BIBentrySTDinterwordspacing

\bibitem{tfrc_347397}
S.~Floyd, M.~Handley, J.~Padhye, and J.~Widmer, ``Equation-based congestion
  control for unicast applications,'' \emph{in Proc. SIGCOMM CCR}, vol.~30, pp.
  43--56, 2000.

\bibitem{draft.rtp.tfrc}
L.~Gharai and C.~Perkins, ``{RTP with TCP Friendly Rate Control},'' March 2011,
  (Work in progress).

\bibitem{saurin:2006:thesis}
A.~Saurin, ``{Congestion Control for Video-conferencing Applications},''
  Master's Thesis, University of Glasgow, December 2006.

\bibitem{VS:Rate}
V.~Singh, J.~Ott, and I.~Curcio, ``Rate adaptation for conversational {3G}
  video,'' in \emph{Proc. of INFOCOM Workshop}, Rio de Janeiro, BR, 2009.

\bibitem{752152}
R.~Rejaie, M.~Handley, and D.~Estrin, ``{Rap: an End-To-End Rate-Based
  Congestion Control Mechanism for Realtime Streams in the Internet},'' in
  \emph{Proc. of INFOCOM}, Mar 1999.

\bibitem{Gharai:2002wt}
L.~Gharai and C.~Perkins, ``{Implementing Congestion Control in the Real
  World},'' in \emph{Proc. of ICME '02}, vol.~1, 2002, pp. 397 -- 400 vol.1.

\bibitem{VladBalan:2007dq}
H.~Vlad~Balan, L.~Eggert, S.~Niccolini, and M.~Brunner, ``{An Experimental
  Evaluation of Voice Quality Over the Datagram Congestion Control Protocol},''
  in \emph{Proc. of IEEE INFOCOM}, 2007.

\bibitem{4397059}
H.~Garudadri, H.~Chung, N.~Srinivasamurthy, and P.~Sagetong, ``{Rate Adaptation
  for Video Telephony in {3G} Networks},'' in \emph{Proc. of PV}, 2007.

\bibitem{VS:HetRate}
V.~Singh, J.~Ott, and I.~Curcio, ``{Rate-control for Conversational Video
  Communication in Heterogeneous Networks},'' in \emph{in Proc. of IEEE WoWMoM
  Workshop}, SFO, CA, USA, 2012.

\bibitem{rmcat-nada}
X.~Zhu and R.~Pan, ``{ NADA: A Unified Congestion Control Scheme for Real-Time
  Media},'' \url{http://tools.ietf.org/html/draft-zhu-rmcat-nada-01}, 2013,
  {IETF Internet Draft}.

\bibitem{rmcat-dflow}
P.~O'Hanlon and K.~Carlberg, ``{Congestion control algorithm for lower latency
  and lower loss media transport},''
  \url{http://tools.ietf.org/html/draft-ohanlon-rmcat-dflow}, 2013, {IETF
  Internet Draft}.

\bibitem{budzisz2011fair}
{\L}.~Budzisz, R.~Stanojevi{\'c}, A.~Schlote, F.~Baker, and R.~Shorten, ``On
  the fair coexistence of loss-and delay-based tcp,'' \emph{IEEE/ACM
  Transactions on Networking (TON)}, vol.~19, no.~6, pp. 1811--1824, 2011.

\bibitem{googleRC}
H.~Alvestrand, S.~Holmer, and H.~Lundin, ``{A Google Congestion Control
  Algorithm for Real-Time Communication on the World Wide Web},'' 2012, {IETF
  Internet Draft}.

\bibitem{rfc5104}
\BIBentryALTinterwordspacing
S.~Wenger, U.~Chandra, M.~Westerlund, and B.~Burman, ``{Codec Control Messages
  in the RTP Audio-Visual Profile with Feedback (AVPF)},'' RFC 5104 (Proposed
  Standard), Internet Engineering Task Force, Feb. 2008. [Online]. Available:
  \url{http://www.ietf.org/rfc/rfc5104.txt}
\BIBentrySTDinterwordspacing

\bibitem{fhm-2013-gcc}
L.~D. Cicco, G.~Carlucci, and S.~Mascolo, ``{Experimental Investigation of the
  Google Congestion Control for Real-Time Flows},'' in \emph{Proc. of ACM
  SIGCOMM 2013 Workshop on Future Human-Centric Multimedia Networking}, 2013.

\bibitem{pv-2013-rrtcc}
V.~Singh, A.~A. Lozano, and J.~Ott, ``{Performance Analysis of Receive-Side
  Real-Time Congestion Control for WebRTC},'' in \emph{Proc. of IEEE Packet
  Video 2013}, 2013.

\bibitem{Zhu:2001tu}
W.~Zhu and Q.~Zhang, ``{Network-Adaptive Rate Control With Unequal Loss
  Protection For Scalable Video Over Internet},'' \emph{in Proc. of Circuits
  and Systems}, 2001.

\bibitem{springerlink:10.1023/A:1022865704606}
Q.~Zhang, W.~Zhu, and Y.-Q. Zhang, ``Network-adaptive rate control and unequal
  loss protection with tcp-friendly protocol for scalable video over
  internet,'' \emph{in Proc. of The Journal of VLSI Signal Processing},
  vol.~34, pp. 67--81, 2003.

\bibitem{871542}
------, ``Network-adaptive rate control with tcp-friendly protocol for multiple
  video objects,'' in \emph{Proc. of ICME}, 2000.

\bibitem{draft-ietf-xrblock-rtcp-xr-discard-rle-metrics-04.txt}
J.~Ott, I.~Curcio, and V.~Singh, ``{RTP Control Protocol (RTCP) Extended
  Reports (XR) for Run Length Encoding (RLE) of Discarded Packets},'' 2011,
  {IETF Internet Draft}.

\bibitem{NgamwongwattanaT10}
B.~Ngamwongwattana and R.~Thompson, ``Sync \& {Sense}: {VoIP} {Measurement
  Methodology for Assessing One-Way Delay Without Clock Synchronization},''
  \emph{in Proc. of IEEE Transactions of Instrumentation and Measurement},
  vol.~59, no.~5, pp. 1318--1326, 2010.

\bibitem{draft.xr.owd}
R.~Brandenburg, K.~Gross, Q.~Wu, F.~Boronat, and M.~Montagud, ``{RTCP XR Report
  Block for One Way Delay metric Reporting},'' 2012, {IETF Internet Draft}.

\bibitem{s4.eval.fw}
{{3GPP S4-080771}}, ``{{MTSI} Video Dynamic Rate Adaptation: Evaluation
  Framework},'' Oct. 2008.

\bibitem{rmcat.req}
R.~Jesup, ``Congestion control requirements for rmcat,'' 2013, {IETF Internet
  Draft}.

\bibitem{ott2012evaluating}
\BIBentryALTinterwordspacing
V.~Singh and J.~Ott, ``Evaluating congestion control for interactive real-time
  media,'' 2013, {IETF Internet Draft}. [Online]. Available:
  \url{http://tools.ietf.org/html/draft-singh-rmcat-cc-eval}
\BIBentrySTDinterwordspacing

\bibitem{rfc5109}
\BIBentryALTinterwordspacing
A.~Li, ``{RTP Payload Format for Generic Forward Error Correction},'' RFC 5109
  (Proposed Standard), Internet Engineering Task Force, Dec. 2007. [Online].
  Available: \url{http://www.ietf.org/rfc/rfc5109.txt}
\BIBentrySTDinterwordspacing

\bibitem{nadu.1070341}
I.~Curcio and D.~Leon, ``Application rate adaptation for mobile streaming,''
  \emph{Proc. of IEEE WOWMOM}, 2005.

\bibitem{4317642}
M.~Wien, H.~Schwarz, and T.~Oelbaum, ``Performance analysis of svc,'' \emph{in
  Proc. of IEEE Transactions on Circuits and Systems for Video Technology},
  vol.~17, no.~9, pp. 1194 --1203, sept. 2007.

\bibitem{Carbone:2010p3502}
M.~Carbone and L.~Rizzo, ``{Dummynet revisited},'' \emph{in Proc. of ACM
  SIGCOMM CCR}, Jan 2010.

\bibitem{perkins12}
\BIBentryALTinterwordspacing
V.~Singh, J.~Ott, and C.~Perkins, ``{Congestion Control for Interactive Media:
  Control Loops \& APIs},'' in \emph{IAB/IRTF Workshop on Congestion Control
  for Interactive Real-Time Communication}, July 2012. [Online]. Available:
  \url{http://csperkins.org/publications/2012/07/iab-cc-workshop.pdf}
\BIBentrySTDinterwordspacing

\bibitem{rfc6051}
\BIBentryALTinterwordspacing
C.~Perkins and T.~Schierl, ``{Rapid Synchronisation of RTP Flows},'' RFC 6051
  (Proposed Standard), Internet Engineering Task Force, Nov. 2010. [Online].
  Available: \url{http://www.ietf.org/rfc/rfc6051.txt}
\BIBentrySTDinterwordspacing

\end{thebibliography}
%\bibliography{IEEEabrv,../rfc}

\end{document}